\documentclass[12pt]{article}
\usepackage{makeidx}
\usepackage{amssymb}
\usepackage{amsfonts}
\usepackage{amsmath}
\usepackage{graphicx}
\usepackage{setspace}
\usepackage{authblk}
\usepackage{units}
\usepackage{appendix}
\usepackage{xparse}
\usepackage{cite}
\usepackage[left=2.1cm,top=2.5cm,right=2.1cm]{geometry}

\pdfoutput=1

\begin{document}

\title{Vacuum instability in time-dependent electric fields. New example of exactly solvable case}
\author{A. I. Breev$^{1}$\thanks{%
breev@mail.tsu.ru}, S. P. Gavrilov$^{1,2}$\thanks{%
gavrilovsergeyp@yahoo.com, gavrilovsp@herzen.spb.ru}, D. M. Gitman$^{1,3,4}$%
\thanks{%
dmitrygitman@hotmail.com}, A. A. Shishmarev$^{1,5}$\thanks{%
a.a.shishmarev@mail.ru}, \\
$^{1}$ Department of Physics, Tomsk State University, \\
Lenin ave. 36, 634050 Tomsk, Russia.\\
$^{2}$ Herzen State Pedagogical University of Russia,\\
Moyka embankment 48, 191186, St. Petersburg, Russia.\\
$^{3}$ P.N. Lebedev Physical Institute, \\
53 Leninskiy ave., 119991 Moscow, Russia.\\
$^{4}$ Institute of Physics, University of S\~{a}o Paulo, \\
Rua do Mat\~{a}o, 1371, CEP 05508-090, S\~{a}o Paulo, SP, Brazil.\\
$^{5}$ Institute of High Current Electronics, SB RAS, \\
Akademichesky ave. 2/3, 634055 Tomsk, Russia.}
\maketitle

\begin{abstract} 
A new exactly solvable case in strong-field quantum electrodynamics with a
time-de\-pen\-dent external electric field is presented. The corresponding
field is given by an analytic function, which is asymmetric (in contrast to
Sauter-like electric field) with respect to the time instant, where it
reaches its maximum value, that is why we call it the analytic asymmetric
electric field. We managed to exactly solve the Dirac equation with such a
field, which made it possible to calculate characteristics of the
corresponding vacuum instability nonperturbatively. We construct the
so-called \textrm{in}- and \textrm{out}-solutions and with their help
calculate mean differential and total numbers of created charged particles,
probability of the vacuum to remain a vacuum, vacuum mean values of current
density and energy-momentum tensor of the particles. We study the vacuum
instability in regimes of rapidly and slowly changing analytic asymmetric
electric field, and compare the obtained results with corresponding ones
obtained earlier for the case of the symmetric Sauter-like electric field.
We also compare exact results in the regime of slowly changing field with
corresponding results obtained within the slowly varying field approximation
recently proposed by two of the authors, thus demonstrating the
effectiveness of such an approximation.

\end{abstract}

\section{Introduction\label{S1}}

Particle creation from the vacuum by strong electromagnetic and
gravitational fields is a remarkable effect (sometimes called the Schwinger
effect \cite{Schwi51}) predicted by quantum field theory ($QFT$). A large
number of articles, reviews and books are devoted to the history of its
theoretical description, possibilities of its observation and applications,
see, e.g., Refs. \cite{Nikis79,BirDav82,GMR85,FraGiS91,Grib94,ruffini} and
references there. $QFT$ with external backgrounds is, to a certain extent,
an appropriate model for theoretical study of the effect. In the framework
of such a model, the particle creation is interpreted as a violation of the
vacuum stability. Backgrounds (external fields) that violate the vacuum
stability are electric-like fields that are able to produce nonzero work
when interacting with charged particles. Creation of charged particles from
the vacuum by electric-like fields needs superstrong field magnitudes
compared with the Schwinger critical field $E_{\mathrm{c}}=m^{2}c^{3}/e\hbar
\simeq 1.3\times 10^{16}\,\mathrm{V}\cdot \mathrm{cm}^{-1}$. Nevertheless,
recent progress in laser physics allows one to hope that this effect will be
experimentally observed in the near future even in laboratory conditions,
see Ref.~\cite{Dun09} for the review. Electron-hole pair creation from the
vacuum (analogue of the electron-positron pair creation) was recently
observed in the graphene by its indirect influence on the graphene
conductivity \cite{Vandecasteele10} (the graphene conductivity modification
due to the particle creation was calculated in \cite{GavGitY12}, some other
relevant effects were discussed in Ref. \cite{Katsnelson}). The need to
consider strong fields in the above mentioned model leads, in turn, to the
need for a nonperturbative consideration of the interaction with external
backgrounds and a development of appropriate methods. Depending on the
structure of such backgrounds, different approaches for calculating the
effect of the vacuum instability in quantum electrodynamics ($QED$) with
strong backgrounds (strong-field $QED$ in what follows) were elaborated. The
most consistent formulation of the particle production problem is formulated
for time-dependent external electric fields that are switched on and off at
infinitely remote times $t\rightarrow \pm \infty $, respectively. A complete
nonperturbative with respect to the external background formulation of
strong-field $QED$ with such external fields was developed in Refs. \cite%
{Gitman,FraGiS91}; it is based on the existence of exact solutions of the
Dirac equation with time dependent external field (more exactly, complete
sets of exact solutions). When such solutions can be found and all the
calculations can be done analytically, we refer to these examples as exactly
solvable cases. Usually considered non-stationary homogeneous electric
fields of constants direction. Electromagnetic vector potentials for such
fields can be chosen as a time-like potential steps step (scalar potentials
being zero), therefore, below we call fields of this type $t$-electric
potential steps, or simply $t$-steps. We note that there are many physically
interesting situations where external backgrounds are constant
(time-independent) but spatially inhomogeneous fields, for example,
concentrated in restricted space areas. A special kind of such backgrounds
are called $x$-electric potential steps (or $x$-steps), in which the field
is inhomogeneous only in one space direction and represents a spatial-like
potential step for charged particles. The $x$-steps can also create
particles from the vacuum, the Klein paradox is closely related to this
process \cite{Klein27,Sauter31a}. Important calculations of the particle
creation by $x$-steps in the framework of the relativistic quantum mechanics
were presented by Nikishov in Refs. \cite{Nikis79,Nikis70a} and later
developed by Hansen and Ravndal in Refs. \cite{HansRavn81,Damour77}. A
general nonperturbative with respect to the external background formulation
of strong-field $QED$ with $x$-steps{\large \ }was developed in Ref. \cite%
{x-case,GavGit20}. Further in this article we discuss problems of vacuum
instability in strong-field $QED$ with $t$-steps only and will not touch
problems related to $x$-steps.

\begin{figure}[ht]
\centering
\includegraphics[width=0.6\textwidth]{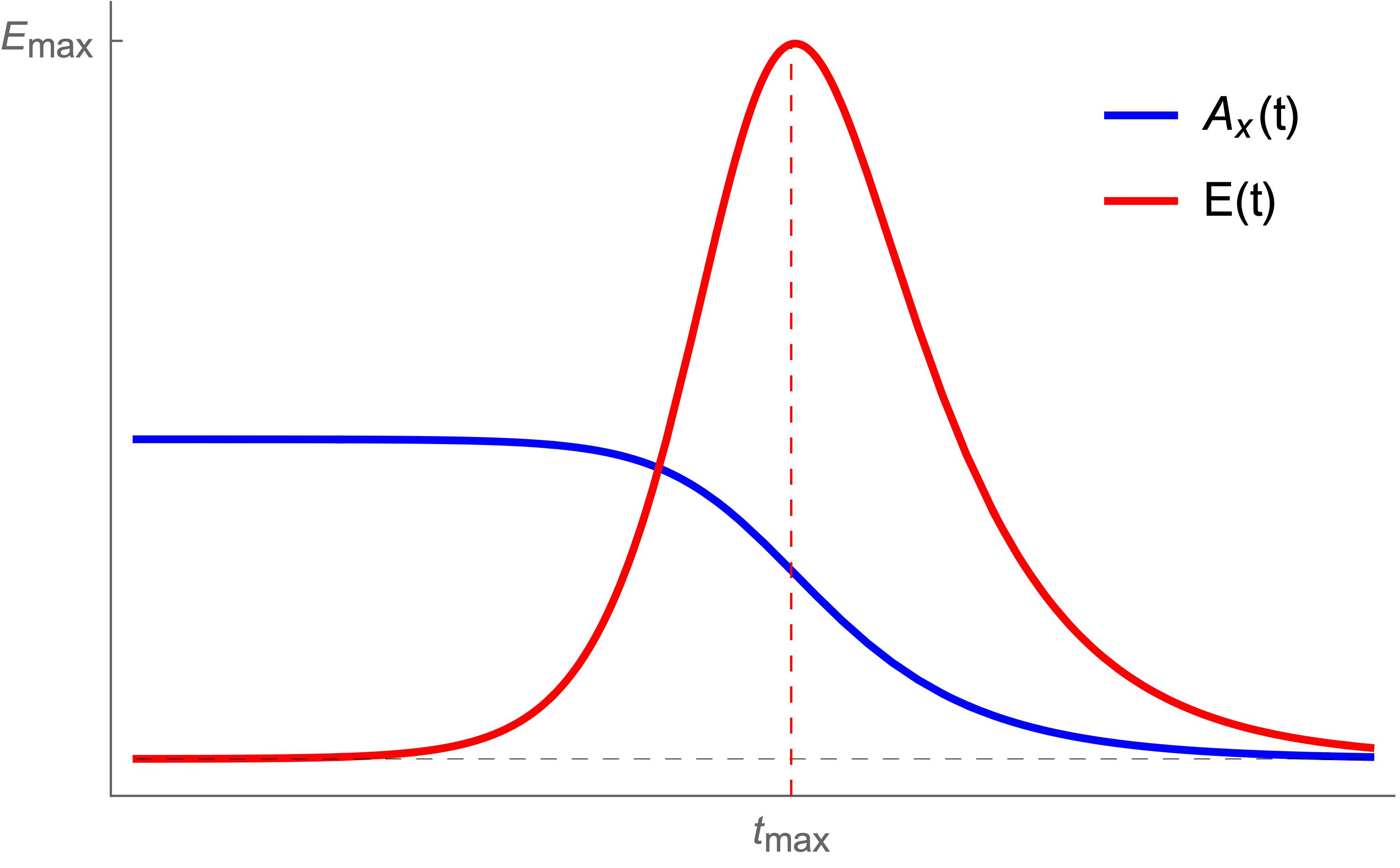}
\caption{General view of electric field (red line) and its vector potential
(blue line) corresponding to a $t$-step.}
\label{fig.0}
\end{figure}

Until now, there are known only few exactly solvable cases in strong-field $%
QED$ with $t$-steps. In this article, we present a new exactly solvable case
in strong-field $QED$ with $t$-steps. Since one of the important goals in
presenting this case, we see its comparison with the already known cases of
this kind; before proceeding to its detailed discussion and studying the
details of the corresponding vacuum instability, we want to briefly recall
the already known exactly solvable cases in strong-field $QED$ with $t$%
-steps. For the generality, the fields are considered in $d=D+1$ -
dimensional Minkowski space-time, parametrized by the coordinates $X=\left(
t,\mathbf{r}\right) $,\ $\mathbf{r}=\left( x^{1}=x,x^{2},\ldots
,x^{D}\right) $. So far, the effect has been considered in homogeneous
fields with constant direction (along one of the axis, usually along the
axis $x$), growing on the interval $\left( -\infty ,t_{\max }\right) $
monotonically from zero to its maximum value $E_{\max }=|\mathbf{E}\left(
t_{\max }\right) |$ at a time instant $t_{\max }$, and then decay
monotonically to zero on the interval $\left( t_{\max },+\infty \right) $.
Their electromagnetic potentials can be chosen as time-like steps, 
\begin{equation}
A^{0}=0,\ \mathbf{A}=\left( A^{1}\left( t\right) ,0,\dots ,0\right) ,\
A^{1}\left( t\right) =A_{x}\left( t\right) =A\left( t\right) ,\ A\left(
-\infty \right) >A\left( +\infty \right),  \label{1.1a}
\end{equation}%
such that 
\begin{equation}
\mathbf{E}\left( t\right) =\left( E^{1}\left( t\right) ,0,\dots ,0\right) ,\
E^{1}\left( t\right) =E_{x}\left( t\right) =E\left( t\right) =-A^{\prime
}\left( t\right) \geq 0,  \label{1.2a}
\end{equation}%
see Fig. \ref{fig.0}. 

\begin{figure}[ht]
\centering
\includegraphics[width=0.8\textwidth]{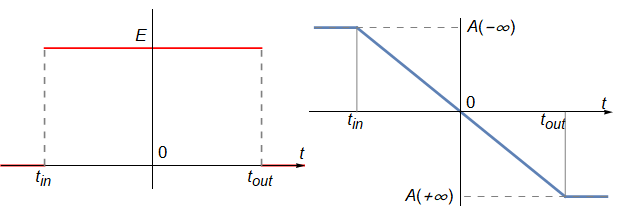}
\caption{$T$-constant electric field and corresponding vector potential.}
\label{fig.2}
\end{figure}

As was already said above there are known only few exactly solvable cases in
strong-field $QED$ with $t$-steps. As the first example of such a case, we
represent the so-called $T$-constant electric field,%
\begin{eqnarray}
A_{x}\left( t\right) &=&\left\{ 
\begin{array}{l}
-Et_{\mathrm{in}},\ \ t\in \left( -\infty ,t_{\mathrm{in}}\right] \\ 
-Et,\ \ t\in \left( t_{\mathrm{in}},t_{\mathrm{out}}\right) \\ 
-Et_{\mathrm{out}},\ \ t\in \left[ t_{\mathrm{out}},+\infty \right)%
\end{array}%
\right. ,  \notag \\
E\left( t\right) &=&\left\{ 
\begin{array}{l}
0,\ \ t\in \left( -\infty ,t_{\mathrm{in}}\right] \\ 
E,\ \ t\in \left( t_{\mathrm{in}},t_{\mathrm{out}}\right) \\ 
0,\ \ t\in \left[ t_{\mathrm{out}},+\infty \right)%
\end{array}%
\right. ,\ \ E>0\ .  \label{1.3a}
\end{eqnarray}%
The $T$-constant field and its corresponding vector potential are displayed
on Fig. \ref{fig.2}. 

The vacuum instability in the $T$-constant electric field was studied in
Refs. \cite{BagGiS75,GavGit96}. It corresponds to a regularized version of
the constant field, in which the electric field remains switched on for all
the time. The vacuum instability in the latter field was studied, e.g., in
Refs. \cite{Schwi51,Nikis79,Nik69}, and in Refs. \cite{Gavri06,GavGi08} by
methods intimately related to the Schwinger effective action.

\begin{figure}[ht]
\centering
\includegraphics[width=0.8\textwidth]{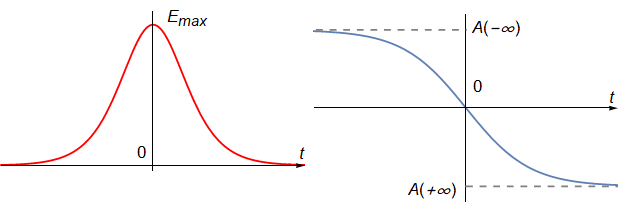}
\caption{Sauter-like electric field and corresponding vector potential.}
\label{fig.1}
\end{figure}

The next example of exactly solvable case is the $t$-step with the so-called
Sauter-like electric field,%
\begin{eqnarray}
&&E\left( t\right) =E_{\max }\cosh ^{-2}\left( t/T_{S}\right) \ ,  \notag \\
&&A_{x}\left( t\right) =-T_{S}\tanh \left( t/T_{S}\right) ,\ E_{\max }>0\ .
\label{1.4}
\end{eqnarray}%
The Sauter-like electric field and its vector potential are displayed on
Fig. \ref{fig.1}. 

The vacuum instability in the Sauter-like electric field was first studied
in Ref. \cite{NarNik70} and then many researchers returned to this problem,
since in the case under consideration it was convenient to test various
approaches, including approximate ones, see, for example, Refs. \cite%
{GelTan16,AdGavGit17} and references therein.

\begin{figure}[ht]
\centering
\includegraphics[width=0.8\textwidth]{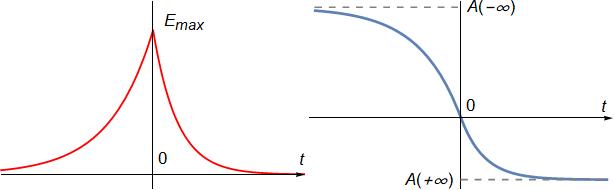}
\caption{Exponential peak field and its vector potential.}
\label{fig.3}
\end{figure}

Next two exactly solvable cases are $t$-steps with the so-called peak
electric fields. The first peak field is an exponentially growing and
decaying electric field. In fact, this peak field is a combination of two
exponential parts, one exponentially increasing and another one
exponentially decreasing,%
\begin{eqnarray}
&&E\left( t\right) =E_{\max }\left\{ 
\begin{array}{l}
\exp \left( k_{1}t\right) ,\ \ t\in \left( -\infty ,0\right] \\ 
\exp \left( -k_{2}t\right) ,\ \ t\in \left( 0,+\infty \right)%
\end{array}%
\right. ,  \notag \\
&&A_{x}\left( t\right) =E_{\max }\left\{ 
\begin{array}{l}
k_{1}^{-1}\left[ -\exp \left( k_{1}t\right) +1\right] ,\ \ t\in \left(
-\infty ,0\right] \\ 
k_{2}^{-1}\left[ \exp \left( -k_{2}t\right) -1\right] ,\ \ t\in \left(
0,+\infty \right)%
\end{array}%
\right. .  \label{1.5}
\end{eqnarray}%
The peak configuration is parametrized by three arbitrary parameters $%
E_{\max }>0,$ $k_{1}>0$\ and$\ k_{2}>0$. We call this field configuration
the exponential peak field. The exponential peak field and its vector
potential are displayed on Fig. \ref{fig.3}. 
The vacuum instability in the exponential peak field was studied in Refs. 
\cite{AdGavGit15,AdGavGit16,GavGitShi17,AdFerGavGit17}.

\begin{figure}[ht]
\centering
\includegraphics[width=0.8\textwidth]{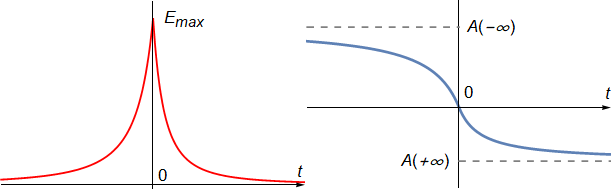}
\caption{Inverse square peak field and its vector potential.}
\label{fig.4}
\end{figure}

The second exactly solvable case of a peak field is also a combination of
two parts, one increasing and another one decreasing, both of them inversely
proportional to square of the time,%
\begin{eqnarray}
&&E\left( t\right) =E_{\max }\left\{ 
\begin{array}{l}
\left( 1-t/\tau _{1}\right) ^{-2},\ \ t\in \left( -\infty ,0\right] \\ 
\left( 1+t/\tau _{2}\right) ^{-2},\ \ t\in \left( 0,+\infty \right)%
\end{array}%
\right. ,\   \notag \\
&&A_{x}\left( t\right) =E_{\max }\left\{ 
\begin{array}{l}
\tau _{1}\left[ 1-\left( 1-t/\tau _{1}\right) ^{-1}\right] ,\ \ t\in \left(
-\infty ,0\right] \\ 
\tau _{2}\left[ \left( 1+t/\tau _{2}\right) ^{-1}-1\right] ,\ \ t\in \left(
0,+\infty \right)%
\end{array}%
\right. .  \label{1.6}
\end{eqnarray}
This peak configuration is parametrized by three arbitrary parameters $%
E_{\max }>0,$ $\tau _{1}>0$\ and$\ \tau _{2}>0$. We call this field
configuration the inverse square peak field. The inverse square peak field
and its vector potential are displayed on Fig. \ref{fig.4}. 
The vacuum instability in the inverse square peak field was studied in Ref. 
\cite{AdGavGit18}.

We note that among the above exactly solvable cases only the Sauter-like
electric field is given by analytic function.

Here we present a new example of exactly solvable case of such type. The
electric field and its potential have the form 
\begin{eqnarray}
&&E\left( t\right) =\frac{E_{0}}{8}\sqrt{1+\exp \left( t/\sigma \right) }%
\cosh ^{-2}\left( t/2\sigma \right) ,\ E_{0}>0,\ \sigma >0\ ,  \notag \\
&&E_{\max }=E\left( t_{\max }\right) =3^{-3/2}E_{0},\ t_{\max }=\sigma \ln
2\ ,  \notag \\
&&A_{x}\left( t\right) =\frac{\sigma E_{0}}{\sqrt{1+\exp \left( t/\sigma
\right) }}\ .  \label{1.7}
\end{eqnarray}

\begin{figure}[ht]
\centering
\includegraphics[width=0.8\textwidth]{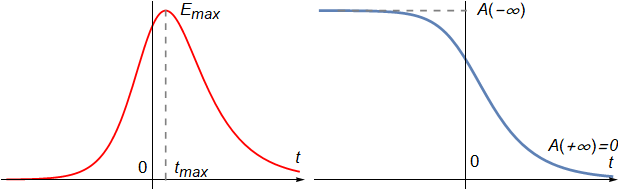}
\caption{Analytic asymmetric field and its potential.}
\label{fig.5}
\end{figure}

In contrast to the Sauter-like electric field this field is asymmetrical
with respect to the time instant $t_{\max }$, where it reaches its maximum
value. We call this field configuration the analytic asymmetric field. The
analytic asymmetric field and its potential are shown of Fig. \ref{fig.5}. 

\begin{figure}[ht]
\centering
\includegraphics[width=0.6\textwidth]{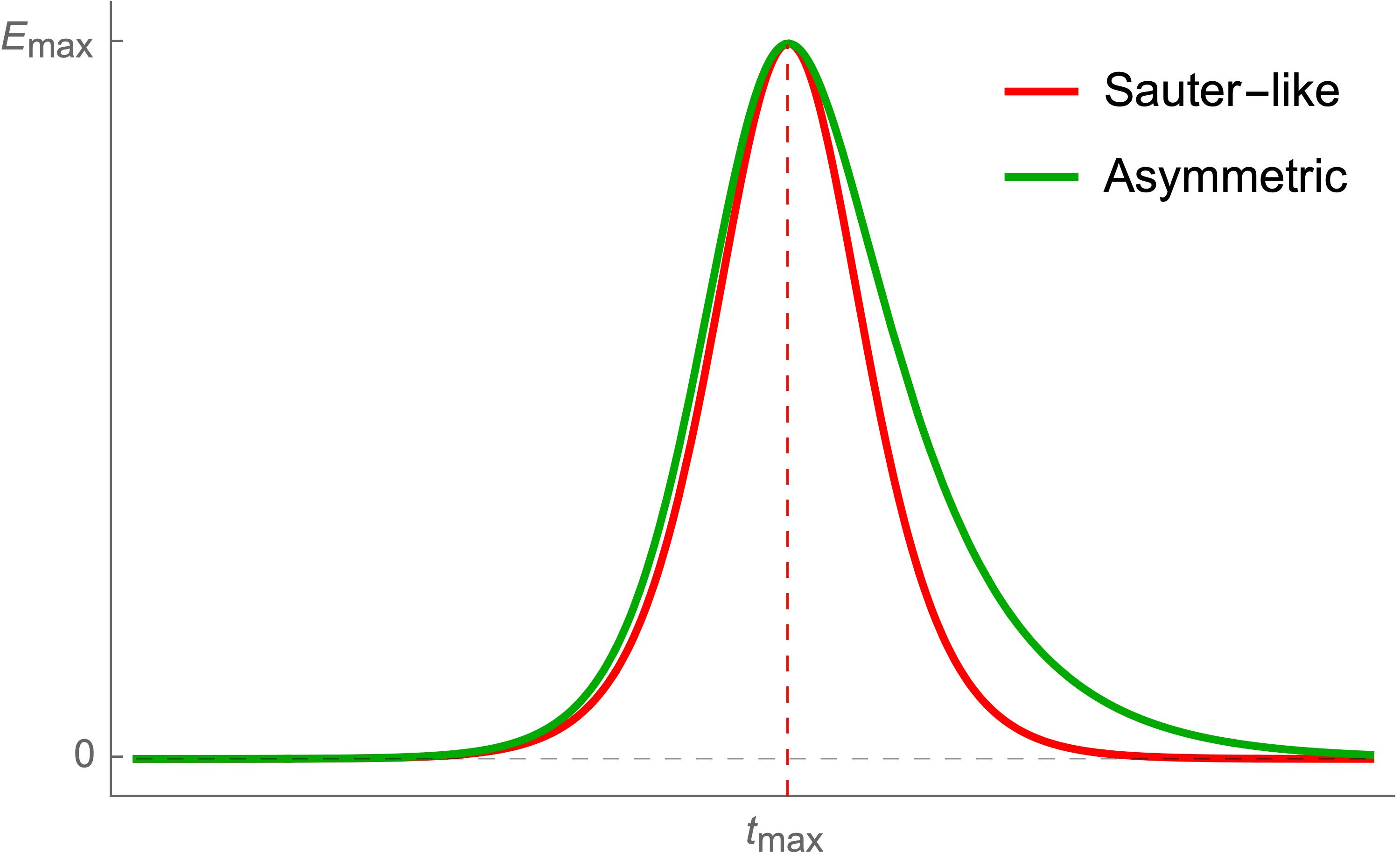}
\caption{Analytic asymmetric field and its potential.}
\label{fig.6}
\end{figure}

It is useful to compare analytic asymmetric electric field and Sauter-like
electric field. On Fig. \ref{fig.6} we present graphs of both fields,
analytic asymmetric field with $\sigma =T_{S}/2$ (by green line) and
Sauter-like field shifted to the right in time by $(T_{S}/2)\log 2$ (by red
line). In this case, both fields reach the same maximum value $E_{\max }$ at
the time instant $t_{\max }=\left( T_{S}/2\right) \log 2 $. 
We note that in the time interval $\left( -\infty ,t_{\max }\right] $ both
fields behave in a very similar way.

We stress that vacuum instability problems which can be analytically studied
using the exactly solvable cases, may be useful in understanding similar
problems in astrophysics, cosmology, and condense matter physics. In
particular, the study of the vacuum instability in the Sauter-like and $T$%
-constant electric fields is instructive for understanding the conductivity
in the graphene and Weyl semimetals as was reported in Refs. \cite%
{GavGitY12,lewkowicz10a,lewkowicz10b,lewkowicz10c,Van+etal10,Zub12,KliMos13,Fil+McL15,VajDorMoe15}%
. Note that the vacuum instability in $T$-constant, exponentially decaying,
and inverse square electric fields has many similarities with the
instability in the de Sitter background, as was noted, e.g., in Refs.~\cite%
{AndMot14,AkhmP15,StStHue16,Ch-Kim20} and cited there works. Besides, using
the exactly solvable cases one can develop and test new approximation
methods for calculating quantum vacuum effects in strong-field $QFT$.

Finally, we would like to note that the Sauter-like, $T$-constant electric
field, and both peak fields are symmetric relative to the point $t=0$, that
plays the role of the time instant $t_{\max }$. In such fields distributions
of created pairs are symmetric with respect of the longitudinal momentum $%
p_{x}$. Almost obvious that the latter symmetry is not inherent in realistic
asymmetric fields. The electric field (\ref{1.7}) considered in this article
is given (similar to Sauter-like field) by an analytical function but is not
symmetric. As we will see below, it corresponds to the exactly solvable case
of $t$-step electric field, thus allowing an analytical and nonperturbative
study of how field asymmetry affects characteristics of the vacuum
instability.

All said above was an incentive for us to study the vacuum instability in
the analytic asymmetric field (\ref{1.7}). This study is the subject of this
article which is organized as follows. In section \ref{S2} we find exact
solutions of the Dirac equation with the analytic asymmetric field (\ref{1.7}%
), in particular, we construct the so-called \textrm{in}- and \textrm{out}%
-solutions which are the basis for calculating characteristics of the vacuum
instability. With their help, in section \ref{S3}, we find nonperturbatively
the vacuum-to-vacuum transition probability as well as differential and
total mean numbers of created pairs. We compare these characteristics with
ones corresponding to other exactly solvable cases. In section \ref{S4} we
study the behavior of obtained physical quantities in the regime of rapidly
and slowly varying analytic asymmetric field (\ref{1.7}). Having in hands an
exact expression for total mean number of created pairs in the regime of
slowly varying field, we compare it with an estimate obtained in an
universal slowly varying field approximation proposed in Ref. \cite{GavGit17}%
, thus demonstrating the effectiveness of the latter. In the same section,
we find mean values of the current density and the energy-momentum tensor of
created particles. Some final remarks are presented in Sect. \ref{S5}.
Useful for us properties of confluent hypergeometric functions are given in
Appendix.

\section{\textrm{In}- and \textrm{out}-solutions\label{S2}}

Let us find solutions to the Dirac equation with electric field (\ref{1.7}).
The Dirac equation in $\left( d=D+1\right) $-dimensional Minkowski
space-time with with such field has the form\footnote{%
We use the relativistic system of units, $\hbar =c=1$.}:%
\begin{equation}
i\partial _{t}\psi \left( X\right) =H\left( t\right) \psi \left( X\right) ,\
\ H\left( t\right) =\gamma ^{0}\left\{ \gamma ^{1}\left[ -i\partial
_{x}+eA_{x}\left( t\right) \right] -i\boldsymbol{\nabla }_{\perp }\gamma
_{\perp }+m\right\} ,  \label{2.4}
\end{equation}%
where $H\left( t\right) $ is a one-particle Dirac Hamiltonian, $\psi \left(
X\right) $ is a $2^{\left[ d/2\right] }$-component spinor ($\left[ d/2\right]
$ stands for the integer part of $d/2$), $e>0$ is the absolute value of the
electron charge, $m$ is the electron mass, and $\gamma ^{\mu }$ are $\gamma $%
-matrices in $d$ dimensions \cite{BraWey35}. The index $\perp $ denotes
components of the momentum operator that are perpendicular to the electric
field.

We seek solutions of Dirac equation in the following form:%
\begin{eqnarray}
&&\psi _{n}\left( X\right) =\exp \left( i\mathbf{pr}\right) \psi _{n}\left(
t\right) ,\;n=\left( \mathbf{p},s\right) \ ,  \notag \\
&&\psi _{n}\left( t\right) =\left\{ \gamma ^{0}i\partial _{t}-\gamma ^{1} 
\left[ p_{x}+eA_{x}\left( t\right) \right] -\boldsymbol{\gamma }\mathbf{p}%
_{\bot }+m\right\} \phi _{n}\left( t\right) \ ,  \label{2.5}
\end{eqnarray}%
where $\psi _{n}\left( t\right) $ and $\phi _{n}\left( t\right) $ are
time-dependent spinors, $n$ is a complete set of quantum numbers
characterizing the solutions. Spin variables can be separated by the
substitution: 
\begin{equation}
\phi _{n}\left( t\right) =\varphi _{n}\left( t\right) v_{\chi ,s},\ \ \chi
=\pm 1,\ \ s=\left( s_{1},s_{2},\dots ,s_{\left[ d/2\right] -1}\right) ,\ \
s_{j}=\pm 1,  \label{2.6}
\end{equation}%
where $\varphi _{n}\left( t\right) $ are some scalar functions and $v_{\chi
,s}$ is a set of constant orthonormalized spinors, satisfying the following
conditions:%
\begin{equation}
\gamma ^{0}\gamma ^{1}v_{\chi ,s}=\chi v_{\chi ,s},\ \ v_{\chi ,s}^{\dag
}v_{\chi ^{\prime },s^{\prime }}=\delta _{\chi ,\chi ^{\prime }}\delta
_{s,s^{\prime }}\ .  \label{2.7}
\end{equation}%
Quantum numbers $s$ and $\chi $ describe the spin polarization (if $d\leq 3$
there are no spin degrees of freedom that are described by the quantum
numbers $s$). The solutions of Dirac equation (\ref{2.5}) which differ only
by values of $\chi $ are linearly dependent, so it is sufficient to work
only with solutions corresponding to one of the values of $\chi $; see Refs. 
\cite{BagGiS75,GavGit96} for more details. The scalar functions $\varphi
_{n}\left( t\right) $ satisfy the following second-order differential
equation:%
\begin{equation}
\left\{ \frac{d^{2}}{dt^{2}}+\left[ p_{x}+eA_{x}\left( t\right) \right]
^{2}+\pi _{\perp }^{2}+i\chi e\dot{A}\left( t\right) \right\} \varphi
_{n}\left( t\right) =0,\ \pi _{\perp }=\sqrt{\mathbf{p}_{\perp }^{2}+m^{2}}\
.  \label{2.8}
\end{equation}

Now, we transform equation (\ref{2.8}) to the Heun equation \cite{DLMF,RovAr}
of a special form. To this end, we use the ansatz%
\begin{eqnarray}
&&\varphi _{n}\left( t\right) =\left( 1+z\right) ^{\alpha _{1}}\left(
1-z\right) ^{\alpha _{2}}u_{n}\left( z\right) ,\ \ z=\sqrt{1+\exp \left(
t/\sigma \right) }\ ,  \notag \\
&&\alpha _{1}=i\tau \sigma \sqrt{\left( p_{x}-eE_{0}\sigma \right) ^{2}+\pi
_{\perp }^{2}},\ \ \alpha _{2}=i\tau \sigma \sqrt{\left( p_{x}+eE_{0}\sigma
\right) ^{2}+\pi _{\perp }^{2}}\ .  \label{2.9}
\end{eqnarray}%
Solutions that differ only by the parameter $\tau $ are also linearly
dependent. For us it is sufficient to work only with $\tau =+1$.
Substituting (\ref{2.9}) into (\ref{2.8}), we obtain the Heun equation for
function $u_{n}\left( z\right) $, 
\begin{eqnarray}
&&\hat{H}_{n}u_{n}\left( z\right) =0,\ \hat{H}_{n}=\frac{d^{2}}{dz^{2}}%
+\left( -\frac{1}{z}+\frac{1+2\alpha _{2}}{z-1}+\frac{1+2\alpha _{1}}{z+1}%
\right) \frac{d}{dz}  \notag \\
&&+\frac{z\left[ \alpha _{3}^{2}-\left( \alpha _{1}-\alpha _{2}\right) ^{2}%
\right] +\left( \alpha _{1}-\alpha _{2}+\alpha _{3}\right) }{z\left(
z-1\right) \left( z+1\right) },\ \alpha _{3}=-2i\chi e\sigma ^{2}E_{0},\
\tau =\pm 1.  \label{2.10}
\end{eqnarray}

Let us represent the functions $u_{n}\left( z\right) $ as follows:%
\begin{equation}
u_{n}\left( z\right) =U_{n}\hat{M}_{n}w_{n}\left( \frac{z+1}{2}\right) ,\ \ 
\hat{M}_{n}=\frac{bz-\alpha _{1}+\alpha _{2}-\alpha _{3}}{\left( a-1\right) b%
}\frac{d}{dz}+1,  \label{2.12}
\end{equation}%
where $U_{n}$ are some constants to be defined below, and $w_{n}$ is a set
of special functions, their properties will be discussed below.

For what follows, it should be noted that the differential operator $\hat{H}%
_{n}$ satisfies the identity:%
\begin{eqnarray}
&&\ \hat{H}_{n}\hat{M}_{n}\equiv \hat{B}_{n}\hat{R}_{n},\ \ \hat{R}_{n}=%
\frac{d^{2}}{dz^{2}}+\left( \frac{2\alpha _{1}}{z+1}+\frac{2\alpha _{2}}{z-1}%
\right) \frac{d}{dz}+\frac{\left( a-1\right) b}{z^{2}-1},  \notag \\
&&\ \hat{B}_{n}=\frac{bz-\left( \alpha _{1}-\alpha _{2}+\alpha _{3}\right) }{%
\left( a-1\right) b}\frac{d}{dz}+\frac{2\left( \alpha _{1}+\alpha
_{2}+1\right) -b}{a-1}  \notag \\
&&\ +\frac{1}{\left( a-1\right) b}\left[ \frac{\alpha _{1}-\alpha
_{2}+\alpha _{3}}{z}+\frac{b-\left( \alpha _{1}-\alpha _{2}+\alpha
_{3}\right) }{z-1}-\frac{b+\left( \alpha _{1}-\alpha _{2}+\alpha _{3}\right) 
}{z+1}\right] ,  \label{2.11}
\end{eqnarray}%
where imaginary parameters $a$ and $b$ have the form: 
\begin{equation}
a=\alpha _{1}+\alpha _{2}-\sqrt{2\left( \alpha _{1}^{2}+\alpha
_{2}^{2}\right) -\alpha _{3}^{2}},\ \ b=\alpha _{1}+\alpha _{2}+\sqrt{%
2\left( \alpha _{1}^{2}+\alpha _{2}^{2}\right) -\alpha _{3}^{2}}.
\label{2.11a}
\end{equation}
\ \ 

Then, we chose the functions $w_{n}$ satisfying the equation:%
\begin{equation}
\hat{R}w_{n}\left( \xi \right) =0,\ \ \hat{R}=\frac{d^{2}}{d\xi ^{2}}+\left( 
\frac{2\alpha _{1}}{\xi }+\frac{2\alpha _{2}}{\xi -1}\right) \frac{d}{d\xi }+%
\frac{\left( a-1\right) b}{\xi \left( \xi -1\right) },\ \ \xi =\frac{z+1}{2},
\label{2.11b}
\end{equation}%
which admits solutions in terms of hypergeometric functions. Taking into
account equation (\ref{2.11b}), it is a trivial matter to show that
functions (\ref{2.12}) obey the initial equation (\ref{2.10}).

Let us find the general solution of the hypergeometric equation (\ref{2.11b}%
). To this end we use two pairs of linearly independent solutions that we
denote as $w_{n,i}\left( \xi \right) $; here additional indices $i=1,\ldots
,4$, are introduced to distinguish between solutions with the same quantum
numbers $n$. Solutions $w_{n,1}\left( \xi \right) $ and $w_{n,2}\left( \xi
\right) $ are: 
\begin{eqnarray}
&&w_{n,1}\left( \xi \right) =\xi ^{a-2\alpha _{1}-1}\left( 1-\xi \right)
^{2\alpha _{1}-a-b+1}  \notag \\
&&\times F\left( 2\alpha _{1}-a+1,2-a;2\alpha _{1}-a-b+2;2-\alpha _{1};1-\xi
^{-1}\right) ,  \notag \\
&&w_{n,2}\left( \xi \right) =\xi ^{1-a}F\left( a-1,a-2\alpha
_{1};a+b-2\alpha _{1};1-\xi ^{-1}\right) .  \label{2.13}
\end{eqnarray}%
Functions $F\left( \alpha ,\beta ;\gamma ;\xi \right) $\footnote{%
These functions are also often denoted as $_{2}F_{1}(\alpha ,\beta ;\gamma
;\xi )$.} are Gaussian hypergeometric functions \cite{Bateman}. Solutions (%
\ref{2.13}) are well-defined in a vicinity of the singular point $\xi =1$
(which corresponds to $t\rightarrow -\infty $). Solutions $w_{n,3}\left( \xi
\right) $ and $w_{n,4}\left( \xi \right) $ are: 
\begin{eqnarray}
&&w_{n,3}\left( \xi \right) =\left( -\xi \right) ^{-b}F\left( b,b-2\alpha
_{1}+1;b-a+2;\xi ^{-1}\right) ,  \notag \\
&&w_{n,4}\left( \xi \right) =\left( -\xi \right) ^{1-a}F\left( a-1,a-2\alpha
_{1};a-b;\xi ^{-1}\right) \ .  \label{2.14}
\end{eqnarray}%
They are well-defined in a vicinity of the singular point $\xi =\infty $\
(which corresponds to $t\rightarrow +\infty $). Using functions (\ref{2.13})
and (\ref{2.14}) we construct four complete sets $\varphi _{n,i}\left(
t\right) $, $i=1,2,3,4$, of the corresponding solutions of equation (\ref%
{2.8}).

Now one can move on to building the so-called \textrm{in}- and \textrm{out}%
-solutions $\psi \left( t\right) $ of the Dirac equation. These solutions
have special asymptotics as $t\rightarrow \pm \infty $ and correspond to
initial or final particles and antiparticles. The functions $\varphi \left(
t\right) $ that correspond to spinors $\psi \left( t\right) $, that are 
\textrm{in}-solutions, are denoted as $_{\ \zeta }\varphi _{n}\left(
t\right) $,\ while functions $\varphi \left( t\right) $ that correspond to
spinors $\psi \left( t\right) $, that are \textrm{out}-solutions, are
denoted as $_{\ }^{\zeta }\varphi _{n}\left( t\right) $. Both sets are
classified by a quantum number $\zeta =\pm $ which labels particles ($\zeta
=+$) and antiparticles ($\zeta =-$). The electric field (\ref{1.7}) vanishes
at $|t|\rightarrow \infty ,$ but its vector potentials are different at $%
t\rightarrow -\infty $ and $t\rightarrow +\infty $, see Eq. (\ref{1.7}) .
The above mentioned solutions $_{\ \zeta }\varphi _{n}\left( t\right) $ and $%
_{\ }^{\zeta }\varphi _{n}\left( t\right) $ have the following asymptotic
behavior,%
\begin{eqnarray}
&&\ ^{\zeta }\varphi _{n}\left( t\right) =\ ^{\zeta }\mathcal{N}\exp \left(
-i\ ^{\zeta }\varepsilon _{n}t\right) ,\ \ ^{\zeta }\varepsilon _{n}=\zeta
\omega _{1},\ \ t\rightarrow +\infty ,  \notag \\
&&\ _{\zeta }\varphi _{n}\left( t\right) =\ _{\zeta }\mathcal{N}\exp \left(
-i\ _{\zeta }\varepsilon _{n}t\right) ,\ \ _{\zeta }\varepsilon _{n}=\zeta
\omega _{2},\ \ t\rightarrow -\infty ,  \notag \\
&&\ \omega _{1}=\sqrt{p_{x}^{2}+\pi _{\perp }^{2}},\ \ \omega _{2}=\sqrt{%
\left( p_{x}+eE_{0}\sigma \right) ^{2}+\pi _{\perp }^{2}},  \label{2.15}
\end{eqnarray}%
where$\ _{\zeta }\mathcal{N}$ and$\ ^{\zeta }\mathcal{N}$ are normalization
constants.

Solutions (\ref{2.9}) with the asymptotic conditions (\ref{2.15}) have the
following form:%
\begin{eqnarray}
&&\ _{+}\varphi _{n}\left( t\right) =\ _{+}\mathcal{N}U_{n,1}\left(
1+z\right) ^{\alpha _{1}}\left( 1-z\right) ^{\alpha _{2}}\hat{M}%
_{n}w_{n,1}\left( \frac{z+1}{2}\right) ,\   \notag \\
&&\ _{-}\varphi _{n}\left( t\right) =\ _{-}\mathcal{N}U_{n,2}\left(
1+z\right) ^{\alpha _{1}}\left( 1-z\right) ^{\alpha _{2}}\hat{M}%
_{n}w_{n,2}\left( \frac{z+1}{2}\right) ,  \notag \\
&&\ ^{+}\varphi _{n}\left( t\right) =\ ^{+}\mathcal{N}U_{n,3}\left(
1+z\right) ^{\alpha _{1}}\left( 1-z\right) ^{\alpha _{2}}\hat{M}%
_{n}w_{n,3}\left( \frac{z+1}{2}\right) ,  \notag \\
&&\ ^{-}\varphi _{n}\left( t\right) =\ ^{-}\mathcal{N}U_{n,4}\left(
1+z\right) ^{\alpha _{1}}\left( 1-z\right) ^{\alpha _{2}}\hat{M}%
_{n}w_{n,4}\left( \frac{z+1}{2}\right) ,  \label{2.18}
\end{eqnarray}%
where the constants $U_{n,i}$, $i=1,2,3,4$, are:%
\begin{eqnarray}
&&U_{n,1}=\frac{2^{1-\alpha _{1}-3\alpha _{2}}e^{i\pi \alpha _{2}}\left(
a-1\right) b}{\left( 2\alpha _{2}-1\right) \left( b-\alpha _{1}+\alpha
_{2}-\alpha _{3}\right) }\ ,  \notag \\
&&U_{n,2}=\frac{2^{\alpha _{2}-\alpha _{1}+2}e^{-i\pi \alpha _{2}}\alpha _{2}%
}{a-\alpha _{1}+\alpha _{2}+\alpha _{3}}\ ,  \notag \\
&&U_{n,3}=\frac{2^{-b}e^{-i\pi \left( \alpha _{2}-b\right) }a-1}{a-b-1}\ , 
\notag \\
&&U_{n,4}=\frac{2^{1-a}e^{-i\pi \left( \alpha _{2}-a\right) }b\left(
a-b\right) }{a\left( b-\alpha _{1}+\alpha _{2}-\alpha _{3}\right) -b\left(
a+\alpha _{1}-\alpha _{2}-\alpha _{3}\right) }\ .  \label{2.19}
\end{eqnarray}

The linear independence of solutions $^{\zeta }\varphi _{n}\left( t\right) $%
\ and$\ _{\zeta }\varphi _{n}\left( t\right) $\ (\ref{2.18}) with different $%
\zeta $\ can be proved as follows:\ using well-known relation (\ref{AdF}),
given in Appendix, one sees that Wronskians of the functions $\varphi $\ are
proportional to Wronskians of the functions $w$, namely,%
\begin{eqnarray}
&&\ W\left( \ _{-}\varphi ,\ _{+}\varphi \right) =W\left( \ _{+}\varphi ,\
_{-}\varphi \right) =\Omega _{n}\left( z\right) W\left(
w_{n,1},w_{n,2}\right) ,  \notag \\
&&\ W\left( \ ^{-}\varphi ,\ ^{+}\varphi \right) =W\left( \ ^{+}\varphi ,\
^{-}\varphi \right) =\Omega _{n}\left( z\right) W\left(
w_{n,3},w_{n,4}\right) ,  \notag \\
&&\ \Omega _{n}\left( z\right) =2^{2\left( \alpha _{1}+\alpha _{2}\right)
}\left( 1+z\right) ^{2\alpha _{1}}\left( 1-z\right) ^{2\alpha _{2}}\frac{%
\left( a-b\right) \alpha _{3}+2\left( \alpha _{1}^{2}-\alpha _{2}^{2}\right) 
}{4\left( a-1\right) b}\ .  \label{2.14a}
\end{eqnarray}%
In the case under consideration we have $W\left( w_{n,1},w_{n,2}\right)
=W\left( w_{n,3},w_{n,4}\right) =0$,\ which implies $W\left( \ _{\zeta
}\varphi ,\ _{-\zeta }\varphi \right) =W\left( \ ^{\zeta }\varphi ,\
^{-\zeta }\varphi \right) =0$ and, thus, proves the linear independence of
the corresponding functions $\varphi $.

Since a second-order ordinary linear equation has two linearly independent
solutions, all solutions with the same quantum numbers $n$ are found and
they form complete sets.

We denote by $\left\{ \ _{\zeta }\psi _{n}\left( t\right) \right\} $ and by $%
\left\{ \ ^{\zeta }\psi _{n}\left( t\right) \right\} $ \textrm{in}- and 
\textrm{out}-solutions of\ Dirac equation (\ref{2.4}) which constructed via $%
_{\ \zeta }\varphi _{n}\left( t\right) $ and $^{\zeta }\varphi _{n}\left(
t\right) $ correspondingly by the help of Eqs. (\ref{2.5}) and (\ref{2.6}).

Using the equal-time inner product (which is time-independent for bispinors
under consideration) 
\begin{equation}
\left( \psi ,\psi \right) =\int d\mathbf{r\ }\psi ^{\dag }\left( X\right)
\psi \left( X\right) ,\ \ d\mathbf{r}=dx^{1}dx^{2}\ldots dx^{D}  \label{2.16}
\end{equation}%
of Dirac bispinors, we easily calculate the normalization constants$\
_{\zeta }\mathcal{N}$ and$\ ^{\zeta }\mathcal{N},$ using explicit forms of
their asymptotics,%
\begin{eqnarray}
&&^{\zeta }\mathcal{N}=\ ^{\zeta }CY,\ \ ^{\zeta }C=\left[ 2\omega
_{1}\left( \omega _{1}-\chi \zeta p_{x}\right) \right] ^{-1/2},\ \
Y=V_{\left( d-1\right) }^{-1/2}\ ,  \notag \\
&&_{\zeta }\mathcal{N}=\ _{\zeta }CY,\ \ _{\zeta }C=\left\{ 2\omega _{2}%
\left[ \omega _{2}-\chi \zeta \left( p_{x}+eE_{0}\sigma \right) \right]
\right\} ^{-1/2}\ .  \notag
\end{eqnarray}%
In doing this, we use the standard volume regularization in which the $%
\mathbf{r}$-integration in Eq. (\ref{2.16}) is over a large spatial box of
the volume $V_{(d-1)}=L_{1}\times \cdots \times L_{D}$ in $D$-dimensional
Euclidean space, in this case, periodic boundary conditions are assumed for
the Dirac bispinors. At the same time, one can see that the \textrm{in}- and 
\textrm{out}-solutions with different quantum numbers $n$ are orthogonal.

One can also see that \textrm{in}-solutions with quantum numbers $n$ are
expressed via \textrm{out}-solutions with the same quantum numbers $n\ .$
Thus,%
\begin{equation}
\ ^{\zeta }\psi _{n}\left( t\right) =\sum_{\zeta ^{\prime }}g_{n}\left(
_{\zeta ^{\prime }}|^{\zeta }\right) \ _{\zeta ^{\prime }}\psi _{n}\left(
t\right) \ .  \label{2.20}
\end{equation}%
Coefficients $g\left( _{\zeta ^{\prime }}|^{\zeta }\right) $ can be found
with the help of the inner product (\ref{2.16}),%
\begin{equation}
\left( \ _{\zeta }\psi _{n},\ ^{\zeta ^{\prime }}\psi _{n^{\prime }}\right)
=g_{n}\left( _{\zeta ^{\prime }}|^{\zeta }\right) \delta _{nn^{\prime }},\ \
g_{n}\left( _{\zeta ^{\prime }}|^{\zeta }\right) =g_{n}\left( ^{\zeta
}|_{\zeta ^{\prime }}\right) ^{\ast },\ \ \sum_{\zeta ^{\prime }}g_{n}\left(
^{\zeta }|_{\zeta ^{\prime }}\right) g_{n}\left( _{\zeta ^{\prime }}|^{\zeta
^{\prime \prime }}\right) =\delta _{\zeta \zeta ^{\prime \prime }}\ .
\label{2.21}
\end{equation}

Equations (\ref{2.20}) and (\ref{2.21}) imply the following decomposition of
the corresponding scalar functions:%
\begin{equation}
\ _{\zeta }\varphi _{n}\left( t\right) =g_{n}\left( ^{+}|_{\zeta }\right) \
^{+}\varphi _{n}\left( t\right) +g_{n}\left( ^{-}|_{\zeta }\right) \
^{-}\varphi _{n}\left( t\right) .  \label{2.22}
\end{equation}

Using the Kummer relations (\ref{cm2}) and (\ref{cm1}) for the
hypergeometric equation \cite{Bateman} and decompositions (\ref{2.22}), we
find the coefficients $g_{n}\left( ^{\zeta }|_{\zeta ^{\prime }}\right) $ to
be:%
\begin{eqnarray}
&&g_{n}\left( ^{+}|_{+}\right) =\frac{\ _{+}\mathcal{N}}{\ ^{+}\mathcal{N}}%
\frac{2^{b-\alpha _{1}-3\alpha _{2}+1}\sin \left( \pi b\right) \Gamma \left(
a-b\right) \Gamma \left( b+1\right) }{\left( b-\alpha _{1}+\alpha
_{2}-\alpha _{3}\right) \sin \left( 2\pi \alpha _{2}\right) \Gamma \left(
2\alpha _{1}-b\right) \Gamma \left( 2\alpha _{2}\right) },  \notag \\
&&g_{n}\left( ^{-}|_{+}\right) =-\,\frac{_{+}\mathcal{N}}{^{-}\mathcal{N}}%
\frac{2^{a-\alpha _{1}-3\alpha _{2}}\left( a-\alpha _{1}+\alpha _{2}+\alpha
_{3}\right) \sin \left( \pi a\right) \Gamma \left( b-a\right) \Gamma \left(
a\right) }{\sin \left( 2\pi \alpha _{2}\right) \Gamma \left( b-2\alpha
_{2}+1\right) \Gamma \left( 2\alpha _{2}\right) },  \notag \\
&&g_{n}\left( ^{+}|_{-}\right) =-\,\frac{_{-}\mathcal{N}}{^{+}\mathcal{N}}%
\frac{2^{b-\alpha _{1}+\alpha _{2}+1}\pi \Gamma \left( a-b\right) }{\left(
a-\alpha _{1}+\alpha _{2}+\alpha _{3}\right) \sin \left( 2\pi \alpha
_{2}\right) \Gamma \left( a\right) \Gamma \left( -2\alpha _{2}\right) \Gamma
\left( a-2\alpha _{1}\right) },  \notag \\
&&g_{n}\left( ^{-}|_{-}\right) =\,-\frac{_{-}\mathcal{N}}{^{-}\mathcal{N}}%
\frac{2^{a-\alpha _{1}+\alpha _{2}}\pi \left( b-\alpha _{1}+\alpha
_{2}-\alpha _{3}\right) \Gamma \left( b-a\right) }{\sin \left( 2\pi \alpha
_{2}\right) \Gamma \left( b+1\right) \Gamma \left( -2\alpha _{2}\right)
\Gamma \left( 1-a+2\alpha _{2}\right) },  \label{2.23}
\end{eqnarray}%
where $\Gamma \left( x\right) $ is the gamma-function.

\section{Vacuum instability characteristics\label{S3}}

Here, using exact solutions that were found above, we already can calculate
characteristics of the vacuum instability in the electric field (\ref{1.7}),
namely the vacuum-to-vacuum transition probability $P_{\mathrm{v}},$
differential $N_{n}$ and total $N$ mean numbers of created pairs. As it
follows from the general formulation of strong--field $QED$ with $t$%
-electric potential steps, all these characteristics are expressed via
coefficients (\ref{2.23}),%
\begin{equation}
P_{\mathrm{v}}=\exp \left[ \sum\limits_{n}\ln \left( 1-N_{n}\right) \right]
,\ \ N_{n}=\left\vert g_{n}\left( _{-}\left\vert ^{+}\right. \right)
\right\vert ^{2},\ \ N=\sum\limits_{n}N_{n}.  \label{3.1}
\end{equation}

First, using Eqs. (\ref{2.23}), we find the differential numbers $N_{n}$\ .
They are:%
\begin{eqnarray}
&&\ N_{n}=\frac{\sinh 2\sigma \pi \left( \omega _{0}+\omega _{1}-\omega
_{2}/2\right) \sinh 2\sigma \pi \left( \omega _{0}-\omega _{1}+\omega
_{2}/2\right) }{\sinh 4\sigma \pi \omega _{1}\sinh 2\sigma \pi \omega _{2}}\
,  \notag \\
&&\ \omega _{0}=\frac{1}{2}\sqrt{\left( p_{x}-eE_{0}\sigma \right) ^{2}+\pi
_{\perp }^{2}}\ ,  \notag \\
&&\ \omega _{1}=\sqrt{p_{x}^{2}+\pi _{\perp }^{2}},\ \ \omega _{2}=\sqrt{%
\left( p_{x}+eE_{0}\sigma \right) ^{2}+\pi _{\perp }^{2}}\ .  \label{3.2}
\end{eqnarray}

To further analysis it is convenient to use the gauge invariant longitudinal
kinetic momentum, $P_{x}\left( t\right) =p_{x}+eA_{x}\left( t\right) $\ and
the increment $\Delta W$ of the longitudinal kinetic momentum, 
\begin{equation}
\Delta W=P_{x}\left( t\rightarrow -\infty \right) -P_{x}\left( t\rightarrow
+\infty \right) \ .  \label{3.0}
\end{equation}%
In the case under consideration we have $\Delta W=eE_{0}\sigma $.

Let us analyze the dependence of the calculated quantities on the parameter $%
\sigma $, which determines the shape of the analytic asymmetric electric
field. First, we consider small values of the parameter $\sigma $, 
\begin{equation}
\sigma \ll \left( eE_{0}\right) ^{-1}\sqrt{p_{x}^{2}+\pi _{\perp }^{2}}\ .
\label{3.3}
\end{equation}%
In this case, the electric field (\ref{1.7}) and its potential change
rapidly, and the electric field is a short pulse corresponding to a small
increment $\Delta W$. As it follows from Eq. (\ref{3.2}), in this case, the
differential mean numbers $N_{n}$ are also small enough for any $p_{x}$ and $%
\pi _{\bot }\ $,%
\begin{equation}
N_{n}=\frac{\left( eE_{0}\sigma \right) ^{2}\pi _{\bot }^{2}}{4\left(
p_{x}^{2}+\pi _{\bot }^{2}\right) ^{2}}\left[ 1+O\left( \frac{eE_{0}\sigma }{%
p_{x}^{2}+\pi _{\bot }^{2}}\right) \right] \ .  \label{3.4}
\end{equation}%
It is the case of a weak external field such a result can be derived in the
frame of perturbation theory with respect to the external field. At small
longitudinal momenta, $p_{x}^{2}\ll \pi _{\perp }^{2}$, expression (\ref{3.4}%
) is reduced to the one%
\begin{equation}
N_{n}\approx \frac{\left( \Delta W\right) ^{2}}{4\pi _{\bot }^{2}},
\label{3.5}
\end{equation}%
which coincides with the result obtained, for example, for a weak pulse of $T
$-constant electric field with the height $\Delta W=eET$\ of a corresponding
step in the same range of longitudinal momenta (see \cite{GavGit96}). Since
the form of $T$-const field is quite different from the one of the analytic
asymmetric electric field (\ref{1.7}), we conclude that in the case of a
small $\Delta W$\ the leading term of the distribution $N_{n}$\ is given by
Eq. (\ref{3.5}) that depends only on $\Delta W$\ and does not depend on the
field configuration.

As it follows from\ a semiclassical consideration, most particles produced
at a time instant $t$ have zero longitudinal kinetic momenta and then
accelerated by a field. Thus, we expect to find maximum of the distribution $%
N_{n}$\ when longitudinal kinetic momenta at time $t_{\max }$, 
\begin{equation}
P_{x}\left( t_{\max }\right) =p_{x}^{\prime },\ \ p_{x}^{\prime
}=p_{x}+eE_{0}\sigma /\sqrt{3},\   \label{3.7}
\end{equation}%
is zero. Because of that in what follows we use value $p_{x}^{\prime }$,
which is best suited for analysis of $N_{n}$.

Parameters $\omega $\ (\ref{3.2}), being written in terms $p_{x}^{\prime }$%
,\ have the form:%
\begin{align}
& \omega _{0}=\frac{1}{2}\sqrt{\left[ p_{x}^{\prime }-eE_{0}\sigma \left(
1+1/\sqrt{3}\right) \right] ^{2}+\pi _{\perp }^{2}},  \notag \\
& \omega _{1}=\sqrt{\left( p_{x}^{\prime }-eE_{0}\sigma /\sqrt{3}\right)
^{2}+\pi _{\perp }^{2}},  \notag \\
& \omega _{2}=\sqrt{\left[ p_{x}^{\prime }+eE_{0}\sigma \left( 1-1/\sqrt{3}%
\right) \right] ^{2}+\pi _{\perp }^{2}}.  \label{3.8}
\end{align}

Let us consider the electric field{\Huge \ }for which $\sigma \rightarrow 0$
and $E_{0}^{-1}\rightarrow 0$, such that the increment $\Delta
W=eE_{0}\sigma $ is a finite quantity. At the same time, we assume that for
sufficiently small $\sigma $\ parameters $\omega _{0}$, $\omega _{1} $ and $%
\omega _{2}$ satisfy the following inequalities: 
\begin{equation}
\Delta W\sigma \ll 1,\ \ \max \left\{ \sigma \omega _{0},\sigma \omega
_{1},\sigma \omega _{2}\right\} \ll 1.  \label{3.9}
\end{equation}%
In this case, one can approximate the mean numbers $N_{n}$ as:%
\begin{equation}
\ N_{n}\approx \frac{\omega _{0}^{2}-\left( \omega _{1}-\omega _{2}/2\right)
^{2}}{2\omega _{2}\omega _{1}}.  \label{3.10}
\end{equation}%
If the increment $\Delta W$ is large enough, $\Delta W\gg \pi _{\perp
},p_{x}^{\prime }$, then one can represent Eq. (\ref{3.10}) as follows:%
\begin{equation}
N_{n}=1+O\left( \max \left[ \frac{p_{x}^{\prime }}{\Delta W},\frac{\pi
_{\perp }^{2}}{\left( \Delta W\right) ^{2}}\right] \right) .  \label{3.11}
\end{equation}%
We see that in this case the differential mean numbers $N_{n}$ reach the
maximum possible value for fermions $N_{n}\approx 1$, in a wide range of\
the momenta $p_{x}^{\prime }$\ and $\pi _{\perp }$. The width of each of
these ranges is only one order less that $\Delta W$. Note that this is a
characteristic feature of short strong pulses with large potential steps,
which can be observed in all exactly solvable cases, see review \cite%
{AdGavGit17}.

In what follows we consider the case of big $\sigma $, 
\begin{equation}
\sigma \gg \left( eE_{0}\right) ^{-1/2}\max \left\{ 1,m^{2}/eE_{0}\right\} ,
\label{3.12}
\end{equation}%
which corresponds to a slowly varying electric field, 
\begin{equation}
2\sigma \pi \left[ \omega _{0}\pm \left( \omega _{1}-\omega _{2}/2\right) %
\right] \gg 1,\ \ 4\sigma \pi \omega _{1}\gg 1,\ \ 4\sigma \pi \omega
_{2}\gg 1.  \label{3.13}
\end{equation}%
In this case the differential mean numbers (\ref{3.2}) can be approximately
presented as: 
\begin{equation}
N_{n}\approx \exp \left[ -\pi \tau \right] ,\ \ \tau =2\sigma \left( 2\omega
_{1}+\omega _{2}-2\omega _{0}\right) .  \label{3.14}
\end{equation}

One can check that the mean numbers $N_{n}$ (\ref{3.2}) are negligibly
small, 
\begin{equation}
N_{n}\ll e^{-\pi m^{2}/eE_{\max }}\,,  \label{3.15}
\end{equation}%
in ranges where initial\ and final longitudinal kinetic momenta $P_{x}\left(
-\infty \right) $\ and $P_{x}\left( +\infty \right) $\ are not large,%
\begin{eqnarray}
&&\left\vert p_{x}^{\prime }-eE_{0}\sigma /\sqrt{3}\right\vert <\sqrt{eE_{0}}%
K,  \notag \\
&&\left\vert p_{x}^{\prime }+eE_{0}\sigma \left( 1-1/\sqrt{3}\right)
\right\vert <\sqrt{eE_{0}}K\ ,  \label{3.16}
\end{eqnarray}%
or are too large, 
\begin{eqnarray}
&&\sqrt{eE_{0}}K<p_{x}^{\prime }-eE_{0}\sigma /\sqrt{3},  \notag \\
&&\sqrt{eE_{0}}K<-\left[ p_{x}^{\prime }+eE_{0}\sigma \left( 1-1/\sqrt{3}%
\right) \right] \ .  \label{3.17}
\end{eqnarray}

Thus, one can conclude that main contributions to the mean numbers $N_{n}$
originate from the range of momentum $p_{x}^{\prime }$ determined by the
double inequality: 
\begin{equation}
\left[ \sqrt{eE_{0}}K-eE_{0}\sigma \left( 1-1/\sqrt{3}\right) \right]
<p_{x}^{\prime }<eE_{0}\sigma /\sqrt{3}-\sqrt{eE_{0}}K.  \label{3.18}
\end{equation}

Let us turn to Eq. (\ref{3.14}). The function $\tau $\ has a minimum at $%
p_{x}^{\prime }=0$, which corresponds to the mean number $N_{n}$ of created
particles by a constant uniform field with $E\left( t\right) =E_{\max }$,
then $\tau$ grows monotonically as both $\left\vert p_{x}^{\prime
}\right\vert $\ and $\pi _{\perp }$\ grow. One can show that mean number $%
N_{n}$\ is exponentially small in the range of large transversal momenta, $%
\pi _{\perp }\gtrsim \sqrt{eE_{0}}K$. Therefore the following range of $\pi
_{\perp }$\ is of interest:%
\begin{equation}
\pi _{\perp }\ll \sqrt{eE_{0}}K.  \label{3.19}
\end{equation}%
Conditions (\ref{3.18}) and (\ref{3.19}) determine a range $\Omega $\ of
momenta, beyond which the distribution $N_{n}$\ is negligible. In this
range, the following approximation of the parameter $\tau $ holds true:%
\begin{eqnarray}
&&N_{n}\approx N_{n}^{\mathrm{as}}=\exp \left( -\pi \tau \right) \ ,  \notag
\\
&&\tau \approx \frac{18\sigma \pi _{\perp }^{2}\left( eE_{0}\sigma \right)
^{2}}{2\sqrt{3}\left( eE_{0}\sigma \right) ^{3}-9\sqrt{3}eE_{0}\sigma
p_{x}^{\prime 2}+9p_{x}^{\prime 3}}\ .  \label{3.20}
\end{eqnarray}%
One can see that in two limiting cases:\ 
\begin{eqnarray}
&&eE_{0}\sigma /\sqrt{3}-p_{x}^{\prime }\rightarrow \sqrt{eE_{0}}K,  \notag
\\
&&p_{x}^{\prime }+eE_{0}\sigma \left( 1-1/\sqrt{3}\right) \rightarrow \sqrt{%
eE_{0}}K,  \label{3.22}
\end{eqnarray}%
either $\omega _{1}$\ or $\omega _{2}$\ reach their maxima and the function $%
\tau $ reaches its maximal values $\tau _{\max }^{+}$ or $\tau _{\max }^{-}$
respectively. One can see that $\tau _{\max }^{\pm }\rightarrow \infty $\
as\ $\sqrt{eE_{0}}\sigma \rightarrow \infty $.

In a wide range of transversal and longitudinal momenta, $\pi _{\perp }\ll
eE_{0}\sigma $, $p_{x}^{\prime }\ll eE_{0}\sigma $, the differential mean
numbers $N_{n}$\ do not depend on the parameter $\sigma $\ and coincide with
ones in the constant electric field $E_{\max }$, which are: 
\begin{equation}
N_{n}\approx N_{n}^{0}=e^{-\pi \tau _{0}}\,,\tau _{0}=\left. \tau
\right\vert _{p_{x}^{\prime }=0}=\lambda =\frac{\pi _{\perp }^{2}}{eE_{\max }%
}\,,  \label{3.23}
\end{equation}%
see Refs. \cite{Nik69,Nikis79}.

The total number of pairs\ created from the vacuum ($N=\sum_{n}N_{n}$)\ by a
uniform electric field, is proportional to the space volume $V_{\left(
d-1\right) }$\ as $N=V_{\left( d-1\right) }\rho .$ One can see that the
number density $\rho $\ has the form: 
\begin{equation}
\rho =\frac{1}{\left( 2\pi \right) ^{d-1}}\sum_{s}\int N_{n}\ d\mathbf{p}\,.
\label{3.24}
\end{equation}%
In deriving Eq. (\ref{3.24}) the sum over all momenta $p$\ was transformed
into an integral. Then the integral in the right hand side of Eq. (\ref{3.24}%
) can be approximated by an integral over a subrange $\Omega $\ (given by
Eqs. (\ref{3.18}) and (\ref{3.19})) that represents a dominant contribution
with respect to the total increment to the number density of created
particles,%
\begin{equation}
\Omega :\rho ^{\mathrm{\Omega }}=\frac{1}{\left( 2\pi \right) ^{d-1}}%
\sum_{s}\int_{\mathbf{p\in }\Omega }N_{n}\ d\mathbf{p\ }.  \label{3.25}
\end{equation}%
This quantity can be calculated using Eq.~(\ref{3.25}) with differential
numbers $N_{n}$\ approximated by Eq.~(\ref{3.20}). In this case, the leading
term $\rho ^{\mathrm{\Omega }}$ is formed over the range given by Eqs.~(\ref%
{3.18}) and (\ref{3.19}). In this approximation, the mean numbers $N_{n}$\
do not depend on the spin polarization parameters $s$. Thus, the summation
over $s$\ produces the factor $J_{\left( d\right) }=2^{\left[ d/2\right] \
-1}$ (the number of spin degrees of freedom), such that:%
\begin{equation}
\rho ^{\mathrm{\Omega }}=\frac{J_{\left( d\right) }}{\left( 2\pi \right)
^{d-1}}\int_{\mathbf{p\in }\Omega }N_{n}\,d\mathbf{p\ }.  \label{3.26}
\end{equation}%
Taking into account\ Eq.~(\ref{3.20}), we approximate integral (\ref{3.26})
as: 
\begin{eqnarray}
&&\rho ^{\mathrm{\Omega }}\approx \frac{J_{\left( d\right) }}{\left( 2\pi
\right) ^{d-1}}\int \left( I_{p_{\bot }}^{+}+I_{p_{\bot }}^{-}\right) \,d%
\mathbf{p}_{\bot }\ ,  \notag \\
&&I_{p_{\bot }}^{+}=\int_{0}^{eE_{0}\sigma /\sqrt{3}-\sqrt{eE_{0}}K}e^{-\pi
\tau }dp_{x}^{\prime }\,,\ \ I_{p_{\bot }}^{-}=\int_{-\left[ eE_{0}\sigma
\left( 1-1/\sqrt{3}\right) -\sqrt{eE_{0}}K\right] }^{0}e^{-\pi \tau
}dp_{x}^{\prime }\ .  \label{3.27}
\end{eqnarray}

To calculate $I_{p_{\bot }}^{+}$\ and $I_{p_{\bot }}^{-}$, it is convenient
to represent $\tau $\ as follows $\tau =\lambda \left( q+1\right) $ and to
pass from the integration over $p_{x}^{\prime }$\ to the integration over a
parameter $q$ (the transition to such a variable provides exponential
decrease of the integrand with increasing $q$, and the expansion of the
pre-exponential factor in powers of $q$ has the form of an asymptotic
series). To this end we have to find $p_{x}^{\prime }$ as a function of $q$
using Eq. (\ref{3.20}). Such a function can be found from the cubic equation%
\begin{equation}
r^{3}-\sqrt{3}r^{2}+\frac{2q}{3\sqrt{3}\left( q+1\right) }=0,\ r=\frac{%
p_{x}^{\prime }}{eE_{0}\sigma }\ .  \label{3.31}
\end{equation}%
Note that when $p_{x}^{\prime }\rightarrow eE_{0}\sigma /\sqrt{3}-\sqrt{%
eE_{0}}K$\ and $p_{x}^{\prime }\rightarrow -\left[ eE_{0}\sigma \left( 1-1/%
\sqrt{3}\right) -\sqrt{eE_{0}}K\right] $, the parameter $\tau $\ reaches the
limiting values $\tau _{\max }^{\pm }=\lambda \left( q_{\max }^{\pm
}+1\right) $ respectively. However, since contributions of the factor $\exp
\left( -\pi \tau \right) $\ to integrals (\ref{3.27}) outside of range (\ref%
{3.18}) are exponentially small, one can extend limits of the integration
over $q$\ to $\pm \infty $. Equation (\ref{3.31}) has three real solutions:%
\begin{align}
& r_{1}=\frac{2}{\sqrt{3}}\cos \frac{\alpha \left( q\right) }{3}+1/\sqrt{3}%
,\ \alpha \left( q\right) =\arccos \left[ \left( q+1\right) ^{-1}\right] , 
\notag \\
& r_{2}=-\frac{2}{\sqrt{3}}\cos \left[ \frac{\alpha \left( q\right) }{3}+%
\frac{\pi }{3}\right] +1/\sqrt{3},  \notag \\
& r_{3}=-\frac{2}{\sqrt{3}}\cos \left[ \frac{\alpha \left( q\right) }{3}-%
\frac{\pi }{3}\right] +1/\sqrt{3};  \label{3.33}
\end{align}%
see, e.g., \cite{Korn}.

Since $0<q<+\infty $, the following inequality holds true $0\leq \alpha
\left( q\right) \leq \pi /2,$ which implies: 
\begin{equation}
0\leq \frac{\alpha \left( q\right) }{3}\leq \frac{\pi }{6},\ \ \frac{\pi }{3}%
\leq \left[ \frac{\alpha \left( q\right) }{3}+\frac{\pi }{3}\right] \leq 
\frac{\pi }{2},\ \ -\frac{\pi }{3}\leq \left[ \frac{\alpha \left( q\right) }{%
3}-\frac{\pi }{3}\right] \leq -\frac{\pi }{6}.  \label{3.36}
\end{equation}%
Then 
\begin{equation}
1\leq r_{1}\leq \sqrt{3},\ \ 0\leq r_{2}\leq \frac{1}{\sqrt{3}},\ \ \left( 
\frac{1}{\sqrt{3}}-1\right) \leq r_{3}\leq 0  \label{3.38}
\end{equation}%
such that solutions $r_{2}$\ and $r_{3}$\ represent the parameter $%
p_{x}^{\prime }$ in the subranges\ $p_{x}^{\prime }\in \left( -eE_{0}\sigma
\left( 1-1/\sqrt{3}\right) ,0\right) \ $and $p_{x}^{\prime }\in \left(
0,eE_{0}\sigma /\sqrt{3}\right) $, respectively.

Thus, the integrals $I_{p_{\bot }}^{+}$ and $I_{p_{\bot }}^{-}$ take the
forms:%
\begin{equation}
I_{p_{\perp }}^{\pm }=\pm \frac{2\Delta W}{3\sqrt{3}}\int_{0}^{+\infty }dq%
\frac{\left( q+1\right) ^{-2}}{\sqrt{1-\left( q+1\right) ^{-2}}}\sin \left[ 
\frac{\alpha \left( q\right) }{3}\pm \frac{\pi }{3}\right] \exp \left[ -\pi
\lambda \left( q+1\right) \right] ,  \label{3.39}
\end{equation}%
where $\Delta W=eE_{0}\sigma $, and their sum can be represented as:%
\begin{equation}
I_{p_{\perp }}^{-}+I_{p_{\perp }}^{+}=\frac{2}{3}\Delta W\int_{0}^{+\infty
}dq\frac{\left( 1+q\right) ^{-2}}{\sqrt{1-\left( 1+q\right) ^{-2}}}\cos 
\frac{\alpha \left( q\right) }{3}\exp \left[ -\pi \lambda \left( q+1\right) %
\right] .  \label{3.40}
\end{equation}

Substituting (\ref{3.40}) into (\ref{3.27}) and integrating over $dp_{\perp
}^{\left( d-2\right) }$, we get:%
\begin{align}
& N\approx V_{\left( d-1\right) }\rho ^{\mathrm{\Omega }},\;\rho ^{\mathrm{%
\Omega }}=\beta \frac{\Delta W}{eE_{\max }}k,\ \ \beta =\frac{J_{\left(
d\right) }\left[ eE_{\max }\right] ^{d/2}}{\left( 2\pi \right) ^{d-1}}\exp %
\left[ -\frac{\pi m^{2}}{eE_{\max }}\right] ,  \notag \\
& k=\frac{2}{3}\int_{0}^{+\infty }dq\left( q^{2}+2q\right) ^{-1/2}\left(
q+1\right) ^{-d/2}\cos \frac{\alpha \left( q\right) }{3}\exp \left[ -\frac{%
\pi m^{2}}{eE_{\max }}q\right] .  \label{3.41}
\end{align}

The corresponding probability $P_{\mathrm{v}}$ of the vacuum to remain a
vacuum reads:%
\begin{equation}
P_{\mathrm{v}}=\exp \left[ -\mu N\right] ,\ \ \mu =\sum_{l=0}^{\infty
}\left( l+1\right) ^{-d/2}\exp \left( -l\frac{\pi m^{2}}{eE_{\max }}\right) .
\label{3.42}
\end{equation}

We see that the number density $\rho ^{\mathrm{\Omega }}$ of created pairs
is proportional to\ the increment{\Large \ }$\Delta W$ of kinetic momentum.
This latter quantity defines the total number of states $\Delta WL/2\pi $
with the longitudinal momenta $p_{x}$, in which particles can be created
(here $L$\ is the length of the system along the axis $x$). Note that this
is typical to any slowly varying field \cite{GavGit17}. The latter property
allows one to compare number densities of created pairs due to various
slowly varying electric fields. Among all exactly solvable cases discussed
above, a special place is occupied by the case of $T$-constant electric
field which is constant within the large time interval $T$. The fields of
other exactly solvable cases decrease to zero with distance from the
corresponding maxima, so one may expect that in these cases the pair
production efficiency will be lower. Thus, it is natural to compare the
number density of created pairs by fields with equal $E_{\max }$\ and
increment $\Delta W$\ with the case of $T$-constant field considered in
detail in Refs. \cite{BagGiS75,GavGit96}. In the latter case we set $E_{\max
}=E$\ and $\Delta W=eET$. The density of created pairs due to $T$-constant
field reads:%
\begin{equation}
\rho _{{\scriptstyle T}}^{\mathrm{\Omega }}=\beta \frac{\Delta W}{eE_{\max }}%
.  \label{g1}
\end{equation}%
It is a linear function of the time duration $T$\ and the quantity $\beta $,
given by Eq. (\ref{3.41}), is the pairs production rate. In the exactly
solvable cases with Sauter-like electric field (\ref{1.4}), in the peak
field configurations of the exponential electric field (\ref{1.5}), and the
inverse-square electric field (\ref{1.6}) the latter quantities are:%
\begin{eqnarray}
&&\ \mathrm{(i)}\;\rho _{{\scriptstyle S}}^{\mathrm{\Omega }}=\rho _{{%
\scriptstyle T}}^{\mathrm{\Omega }}k_{{\scriptstyle S}}\;\mathrm{for\;Sauter}%
\text{\textrm{-}}\mathrm{like\ field},  \notag \\
&&\ \mathrm{(ii)}\;\rho _{\mathrm{p}}^{\mathrm{\Omega }}=\rho _{{%
\scriptstyle T}}^{\mathrm{\Omega }}k_{\mathrm{p}}\;\mathrm{%
for\;exponential\;peak\ field},  \notag \\
&&\ \mathrm{(iii)}\;\rho _{\mathrm{sq}}^{\mathrm{\Omega }}=\rho _{{%
\scriptstyle T}}^{\mathrm{\Omega }}k_{\mathrm{sq}}\;\mathrm{%
for\;inverse\;square\;peak\ field},  \label{g2}
\end{eqnarray}%
where%
\begin{eqnarray}
&&k_{{\scriptstyle S}}=\frac{1}{2}\int_{0}^{\infty }dqq^{-1/2}(q+1)^{-\left(
d+1\right) /2}\exp \left( -q\pi \frac{m^{2}}{eE_{\max }}\right) ,  \notag \\
&&k_{\mathrm{p}}=\int_{0}^{\infty }\frac{dq}{(q+1)^{d/2+1}}\exp \left( -q\pi 
\frac{m^{2}}{eE_{\max }}\right) ,  \notag \\
&&k_{\mathrm{sq}}=\frac{1}{2}\int_{0}^{\infty }\frac{dq}{(q+1)^{d/2}}\exp
\left( -q\pi \frac{m^{2}}{eE_{\max }}\right) .  \label{g3a}
\end{eqnarray}

Formulas (\ref{g3a}) and (\ref{3.41}) show how differences in the shapes of $%
t$-steps affect the integrands for factors $k$'s. These factors for a not so
strong electric field ($m^{2}/eE_{\max }>$\ $1$) can be approximated as: 
\begin{equation}
k\approx \frac{\sqrt{2}}{3}k_{{\scriptstyle S}},\;k_{{\scriptstyle S}%
}\approx \frac{\sqrt{eE_{\max }}}{m},\;k_{\mathrm{sq}}\approx \frac{1}{2}k_{%
\mathrm{p}},\;k_{\mathrm{p}}\approx \frac{eE_{\max }}{\pi m^{2}}\ .
\label{g4}
\end{equation}%
In this case, we see that the number density of created pairs by analytic
asymmetric and Sauter-like fields comparable to each other but less the
density $\rho _{{\scriptstyle T}}^{\mathrm{\Omega }}$ in $T$-constant field
by the order factor\ $\sqrt{eE_{\max }}/m$. The mean number densities $\rho
_{\mathrm{p}}^{\mathrm{\Omega }}$ and $\rho _{\mathrm{sq}}^{\mathrm{\Omega }%
} $ are comparable to each other, but due to the factor $eE_{\max }/m^{2}$
are less than the density $\rho _{{\scriptstyle T}}^{\mathrm{\Omega }}$\ .
The obtained estimates mean that for the case of not very strong electric
fields, vacuum instability effects by the analytic asymmetric and
Sauter-like fields decrease over time, deviating from their maximum values,
much more slowly than by sharp-peak fields.

In the case of a strong electric field ($\pi m^{2}/eE_{\max }<$\ $1$)\
exponential factors in integrals (\ref{g3a}) can be approximated by units.
Then, in this approximation, the factors $k$'s differ slightly and have the
form:%
\begin{eqnarray}
&&k\approx \left\{ 
\begin{array}{l}
{0.65\;\mathrm{if\;}d=4} \\ 
{0.77\;\mathrm{if\;}d=3}%
\end{array}%
\right. \! ,\ k_{{\scriptstyle S}}\approx \left\{ 
\begin{array}{l}
{0.67\;\mathrm{if\;}d=4} \\ 
{0.79\;\mathrm{if\;}d=3}%
\end{array}%
\right. \! , \;k_{\mathrm{sq}}\approx \frac{1}{d-1},\;k_{\mathrm{p}}\approx 
\frac{2}{d}\ .  \label{g5}
\end{eqnarray}%
Thus, if all the above mentioned $t$-steps are strong enough, intensities of
the corresponding pair productions are quite similar.

\section{Slowly varying field approximation\label{S4}}

A new semiclassical approximation approach that is not restricted by a
smallness of differential mean numbers of created pairs was recently
proposed for treating the vacuum instability in strong-field $QED$ with $t$%
-steps slowly varying with time \cite{GavGit17}. This approach is closely
related to the leading term approximation of derivative expansion in
field-theoretic calculations \cite{DunHal98,GusSho96,GusSho99} (see Ref.~%
\cite{Dunn04} for a review). In fact, it is an extension of a locally
constant field approximation (LCFA) for calculating vacuum mean values of
physical quantities. It maintains the nonperturbative character of $QED$
calculations even in the absence of the exact solutions. In this
approximation one can see an universal character of the vacuum effects
caused by a strong electric field, defining the slowly varying regime in
general terms. In particular, one finds\ representations for the total
density of created pairs and vacuum mean values of the current density and
energy-momentum tensor as a functional of an external electric field. In
this section we compare the results of such an approximation to ones
elaborated from exact solutions presented above for sufficiently large $%
\sigma $.

We call $E\left( t\right) $ a slowly varying electric field on a time
interval $\Delta t$ from $t$ to $t+\Delta t$ if the following condition
holds true:%
\begin{equation}
\left\vert \frac{\overline{\dot{E}\left( t\right) }\Delta t}{\overline{%
E\left( t\right) }}\right\vert \ll 1,  \label{svf1}
\end{equation}%
where $\overline{E\left( t\right) }$ and $\overline{\dot{E}\left( t\right) }$
are mean values of $E\left( t\right) $ and $\dot{E}\left( t\right) $ on the
time interval $\Delta t$, respectively, and $\Delta t$ is significantly
larger than the time scale $\Delta t_{\mathrm{sc}}$,%
\begin{eqnarray}
&&\Delta t/\Delta t_{\mathrm{sc}}\gg 1,  \notag \\
&&\Delta t_{\mathrm{sc}}=\left[ e\overline{E\left( t\right) }\right]
^{-1/2}\max \left\{ 1,m^{2}/e\overline{E\left( t\right) }\right\} .
\label{svf2}
\end{eqnarray}%
Property (\ref{svf1}) is inherent to field (\ref{1.7}) for sufficiently
large $\sigma $ satisfying condition (\ref{3.12}) and for $\Delta t$
satisfying as well as Eq. (\ref{svf2}) and the condition 
\begin{equation}
\Delta t/\sigma \ll 1.  \label{svf3}
\end{equation}%
In this case one can approximate the mean value $\overline{E\left( t\right) }
$ in the time interval $\Delta t$ as $\overline{E\left( t\right) }\approx
E\left( t\right) $. For a given $p_{\bot }$ one can consider the time
interval $\Delta t$ as a sufficiently large if%
\begin{equation}
\Delta t\sqrt{eE\left( t\right) }\gg \max \left\{ 1,\lambda \left( t\right)
\right\} ,\;\lambda \left( t\right) =\pi _{\perp }^{2}/eE\left( t\right) .
\label{svf4}
\end{equation}

Since the field under consideration (as all the above mentioned fields)
weakens as $t\rightarrow \pm \infty $ there always exist some time instants $%
t_{\mathrm{in}}$ and $t_{\mathrm{out}}$ such that for any $\Delta t,$ which
satisfies condition (\ref{svf3}),\ the parameter $\lambda \left( t\right) $
achieves critical values $\lambda _{\mathrm{out/in}}=\Delta t\sqrt{eE\left(
t_{\mathrm{out/in}}\right) }>1$ respectively. For big $\lambda \left(
t\right) ,$ satisfying inequalities $\lambda \left( t\right) >$ $\lambda _{%
\mathrm{out}}$ or $\lambda \left( t\right) >$ $\lambda _{\mathrm{in}}$ ,
condition (\ref{svf4}) is not valid. Because of this the slowly varying
field approximation is applicable only in the domain of a strong enough
field when $t_{\mathrm{in}}<t<t_{\mathrm{out}}\ $. However, the violation of
the vacuum stability by a small electric field, $E\left( t\right) <E\left(
t_{\mathrm{out/in}}\right) $, is negligibly small.

In the domain of a strong enough field the leading term in the density $\rho
^{\mathrm{\Omega }}$ in the slowly varying field approximation reads:%
\begin{equation}
\rho ^{\mathrm{\Omega }}\approx \frac{J_{\left( d\right) }}{\left( 2\pi
\right) ^{d-1}}\int_{t_{\mathrm{in}}}^{t_{\mathrm{out}}}eE\left( t\right)
dt\int d\mathbf{p}_{\bot }N_{n}^{\mathrm{univ}},\ \ N_{n}^{\mathrm{univ}%
}=\exp \left[ -\pi \frac{\pi _{\bot }^{2}}{eE\left( t\right) }\right] ,
\label{uni2}
\end{equation}%
and the probability of the vacuum to remain a vacuum has the form:%
\begin{equation}
P_{v}\approx \exp \left\{ -\frac{V_{\left( d-1\right) }J_{\left( d\right) }}{%
\left( 2\pi \right) ^{d-1}}\sum_{l=0}^{\infty }\int_{t_{\mathrm{in}}}^{t_{%
\mathrm{out}}}dt\left( -1\right) ^{\left( 1-\kappa \right) l/2}\frac{\left[
eE\left( t\right) \right] ^{d/2}}{\left( l+1\right) ^{d/2}}\exp \left[ -\pi 
\frac{\left( l+1\right) m^{2}}{eE\left( t\right) }\right] \right\} ,
\label{uni5}
\end{equation}%
respectively, with the field $E\left( t\right) $ given by Eq. (\ref{1.7}).
However, the contribution due to a small field, $E\left( t\right) <E\left(
t_{\mathrm{out/in}}\right) $, to the effect of pair production is negligibly
small. Because of this, we can extend the integration limits as $t_{\mathrm{%
in}}\rightarrow -\infty $ and $t_{\mathrm{out}}\rightarrow +\infty $.

Let us compare these results with the ones presented in Eq. (\ref{3.39}).
One can represent Eq. (\ref{uni2}) in the form%
\begin{eqnarray}
&&\rho ^{\mathrm{\Omega }}\approx \frac{J_{\left( d\right) }}{\left( 2\pi
\right) ^{d-1}}\int d\mathbf{p}_{\bot }\left( J_{p_{\bot }}^{+}+J_{p_{\bot
}}^{-}\right) \,,  \notag \\
&&J_{p_{\bot }}^{+}=\int_{t_{\max }}^{\infty }dt\left[ eE\left( t\right) %
\right] N_{n}^{\mathrm{univ}},\;J_{p_{\bot }}^{-}=\int_{-\infty }^{t_{\max
}}dt\left[ eE\left( t\right) \right] N_{n}^{\mathrm{univ}}\ .  \label{gav1}
\end{eqnarray}%
The electric field $E\left( t\right) $, given by Eq. (\ref{1.7}) is related
to the quantity $W\left( t\right) $ as 
\begin{equation*}
W\left( t\right) =\frac{-eE_{0}\sigma }{\sqrt{1+\exp \left( t/\sigma \right) 
}},
\end{equation*}%
such that $eE\left( t\right) dt=dW\left( t\right) $. On the other hand, one
can relate the functions $E\left( t\right) $ and $W\left( t\right) $ via a
cubic equation,%
\begin{equation}
y^{3}-y-2E\left( t\right) /E_{0}=0,  \label{gav2}
\end{equation}%
where the notation $y=W\left( t\right) /\left( eE_{0}\sigma \right) $ is
used. It is convenient to introduce a variable $q$ such that%
\begin{equation}
q=\left( 3\sqrt{3}E\left( t\right) /E_{0}\right) ^{-1}-1.  \label{gav3}
\end{equation}%
We can express $W\left( t\right) $ as a function of the field $E\left(
t\right) $ or as a function of the variable $q$ using solutions of equation (%
\ref{gav2}). This equation has three real solutions,%
\begin{align}
& y_{1}=\frac{2}{\sqrt{3}}\cos \frac{\alpha \left( q\right) }{3},\ y_{2}=-%
\frac{2}{\sqrt{3}}\cos \left[ \frac{\alpha \left( q\right) }{3}+\frac{\pi }{3%
}\right] ,  \notag \\
& y_{3}=-\frac{2}{\sqrt{3}}\cos \left[ \frac{\alpha \left( q\right) }{3}-%
\frac{\pi }{3}\right] ,\ \alpha \left( q\right) =\arccos \left[ \left(
q+1\right) ^{-1}\right] ,  \label{gav4}
\end{align}%
see, e.g., \cite{Korn}. Since $W\left( t\right) $ is negative, only the
solutions $y_{2,3}$ are relevant. One can see that for solutions $y_{2,3}$
the differential $dW\left( t\right) $ takes the form:%
\begin{equation}
dW\left( t\right) =\frac{2eE_{0}\sigma }{3\sqrt{3}}\sin \left[ \frac{\alpha
\left( q\right) }{3}\pm \frac{\pi }{3}\right] \frac{\left( 1+q\right) ^{-1}dq%
}{\sqrt{q^{2}+2q}}.  \label{gav5}
\end{equation}%
Passing from the integration over $t$ to the integration over the parameter $%
q$ in Eq. (\ref{gav1}), we find:%
\begin{equation}
J_{p_{\bot }}^{\pm }=I_{p_{\bot }}^{\pm }\ ,  \label{gav6}
\end{equation}%
where the quantities $I_{p_{\bot }}^{\pm }$ are given by Eq. (\ref{3.39}).

It follows from Eq. (\ref{gav6}) that the density of created pairs (\ref%
{uni2}) and the probability of the vacuum to remain a vacuum (\ref{uni5})
obtained with the help of the slowly varying field approximation coincide
with expressions (\ref{3.41}) and (\ref{3.42}), respectively.

The slowly varying field approximation also reproduces leading terms for the
mean values (calculated exactly in Ref. \cite{GavGit17}) 
\begin{eqnarray}
&&\ \langle j^{\mu }\left( t\right) \rangle =\langle 0,\mathrm{in}|j^{\mu
}|0,\mathrm{in}\rangle ,\ \ \langle T_{\mu \nu }\left( t\right) \rangle
=\langle 0,\mathrm{in}|T_{\mu \nu }|0,\mathrm{in}\rangle \,,  \notag \\
&&\ j^{\mu }=\frac{-e}{2}\left[ \overline{\hat{\Psi}}\left( x\right) ,\gamma
^{\mu }\hat{\Psi}\left( x\right) \right] \,,\ \ T_{\mu \nu }=\frac{1}{2}%
\left( T_{\mu \nu }^{\mathrm{can}}+T_{\nu \mu }^{\mathrm{can}}\right) \,, 
\notag \\
&&\ T_{\mu \nu }^{\mathrm{can}}=\frac{1}{4}\left\{ \left[ \overline{\hat{\Psi%
}}\left( x\right) ,\gamma _{\mu }P_{\nu }\hat{\Psi}\left( x\right) \right] +%
\left[ P_{\nu }^{\ast }\overline{\hat{\Psi}}\left( x\right) ,\gamma _{\mu }%
\hat{\Psi}\left( x\right) \right] \right\} \,,  \notag \\
&&\ P_{\mu }=i\partial _{\mu }+eA_{\mu }\left( x\right) ,\ \overline{\hat{%
\Psi}}\left( x\right) =\hat{\Psi}^{\dagger }\left( x\right) \gamma ^{0}\ ,
\label{xy}
\end{eqnarray}%
of the current density and of the energy-momentum tensor ($EMT$) operators $%
j^{\mu }$ and $T_{\mu \nu }$ respectively. Here $\hat{\Psi}\left( x\right) $
is the operator of the quantum Dirac field. These terms read:%
\begin{eqnarray}
&\langle j^{1}\left( t_{\mathrm{out}}\right) \rangle \approx &2e\rho ^{%
\mathrm{\Omega }},\;\rho ^{\mathrm{\Omega }}=\frac{J_{(d)}}{(2\pi )^{d-1}}%
\int_{t_{\mathrm{in}}}^{t_{\mathrm{out}}}\left[ eE\left( t\right) \right]
^{d/2}\exp \left[ -\pi \frac{m^{2}}{eE\left( t\right) }\right] dt,  \notag \\
&\langle T_{00}\left( t_{\mathrm{out}}\right) \rangle \approx &\langle
T_{11}\left( t_{\mathrm{out}}\right) \rangle \approx \frac{J_{(d)}}{(2\pi
)^{d-1}}\int_{t_{\mathrm{in}}}^{t_{\mathrm{out}}}\left[ W\left( t_{\mathrm{%
out}}\right) -W\left( t\right) \right] \left[ eE\left( t\right) \right]
^{d/2}\exp \left[ -\frac{\pi m^{2}}{eE\left( t\right) }\right] dt,  \notag \\
&\langle T_{ll}\left( t_{\mathrm{out}}\right) \rangle \approx &\frac{J_{(d)}%
}{(2\pi )^{d}}\int_{t_{\mathrm{in}}}^{t_{\mathrm{out}}}\frac{\left[ eE\left(
t\right) \right] ^{d/2+1}}{\left[ W\left( t_{\mathrm{out}}\right) -W\left(
t\right) \right] }\exp \left[ -\frac{\pi m^{2}}{eE\left( t\right) }\right]
dt,\;l=2,...,D\ .  \label{uni11}
\end{eqnarray}

As before, we can assume $t_{\mathrm{in}}\rightarrow -\infty $ and $t_{%
\mathrm{out}}\rightarrow \infty $. We note that $\rho ^{\mathrm{\Omega }}$\
defined by Eq. (\ref{uni11}) can be equivalently represented by Eq. (\ref%
{3.41}). Using the change of variables (\ref{gav4}) and (\ref{gav5})
together with Eq. (\ref{gav6}), we obtain the following result: 
\begin{eqnarray}
&&\ \langle T_{00}\left( t_{\mathrm{out}}\right) \rangle \approx \langle
T_{11}\left( t_{\mathrm{out}}\right) \rangle \approx \frac{J_{(d)}}{(2\pi
)^{d-1}}\frac{2\left( \Delta W\right) ^{2}}{3\sqrt{3}}\left( eE_{\max
}\right) ^{d/2-1}  \notag \\
&&\times \int_{0}^{+\infty }\frac{\left( 1+q\right) ^{-d/2}}{\sqrt{q^{2}+2q}}%
\cos \left[ \frac{2\alpha \left( q\right) }{3}\right] \exp \left[ -\frac{\pi
m^{2}}{eE_{\max }}\left( q+1\right) \right] dq\ ,  \notag \\
&&\ \langle T_{ll}\left( t_{\mathrm{out}}\right) \rangle \approx \frac{%
J_{(d)}}{(2\pi )^{d-1}}\frac{\left( eE_{\max }\right) ^{d/2}}{\sqrt{3}}%
\int_{0}^{+\infty }\frac{\left( 1+q\right) ^{-d/2-1}}{\sqrt{q^{2}+2q}} 
\notag \\
&&\ \times \left[ \cos \frac{2\alpha \left( q\right) }{3}-\frac{1}{2}\right]
^{-1}\exp \left[ -\frac{\pi m^{2}}{eE_{\max }}\left( q+1\right) \right] dq,\
l=2,...,D\ .  \label{gavEMTa}
\end{eqnarray}

As was demonstrated above, the number density of created pairs $\rho ^{%
\mathrm{\Omega }}$ given by Eq. (\ref{3.41}) is proportional to the
increment $\Delta W$. The mean values $\langle T_{00}\left( t_{\mathrm{out}%
}\right) \rangle \approx \langle T_{11}\left( t_{\mathrm{out}}\right)
\rangle $ are proportional to the square of the increment,\ while the mean
values $\langle T_{ll}\left( t_{\mathrm{out}}\right) \rangle $\ are
independent on $\Delta W$ if the latter is sufficiently large. Note that
such a behavior of the mean values is typical for slowly varying fields \cite%
{GavGit17}. This observation allows one to compare effects of the vacuum
instability due to various electric fields.

\section{Final remarks\label{S5}}

As stated in the Introduction, until now, there were known only few exactly
solvable cases in strong-field $QED$ with $t$-steps. In the paper we present
a new case of this kind, the corresponding $t$-step is given by a
time-dependent analytic asymmetric field (\ref{1.7}). For a nonperturbative
analysis of the vacuum instability generated by such an external field, we
have followed the well-known general approach proposed in the works \cite%
{Gitman,FraGiS91}, based on the use of the corresponding exact solutions (in
particular \textrm{in}- and \textrm{out}-solutions) of the Dirac equation.

One of the main and new result of the work was finding such solutions. It
must be said that this problem turned out to be completely non-trivial.
Unlike previously mentioned exactly solvable cases, for the asymmetric
analytical field we had to apply an original method based on an analogue of
Darboux transformation. With help of this method, we find solutions of the
Dirac equation in the form of the differential transformation (\ref{2.12})
of the Gaussian hypergeometric functions. Then we construct complete sets of 
\textrm{in}- and \textrm{out}-solutions of the Dirac equation with the
asymmetric analytic field, see Sec. \ref{S2}.

With the help of these sets (following the above mentioned nonperturbative
technics), we have calculated exactly basic characteristics of the vacuum
instability in the electric field under consideration, namely the
vacuum-to-vacuum transition probability $P_{\mathrm{v}},$ differential $%
N_{n} $ and total $N$ mean numbers of created pairs, see Sec. \ref{S3}.
Next, we compare the obtained characteristics with the corresponding
characteristics of vacuum instability in other exactly solvable cases. We
analyze the dependence of the calculated quantities on the time scale
parameter $\sigma $, which determines the shape of the analytic asymmetric
electric field.

We note that in the case of a weak analytic asymmetric field, the obtained
results are reduced to results which can be derived in the framework of a
perturbation theory with respect to the external field. This, in particular,
is evidenced by the expression for the differential mean numbers (\ref{3.5})
obtained for the case of the weak field.

If the analytic asymmetric electric field is strong and the increment $%
\Delta W$ of the longitudinal momentum is large enough we deal with the case
of a rapidly changing electric field. In this case, as it follows from Eq. (%
\ref{3.11}), the differential mean numbers $N_{n}$ reach their maximum
possible for fermions values $N_{n}\approx 1$, in wide ranges of the momenta 
$p_{x}^{\prime }$\ and\ $\pi _{\perp }$. The width of each of these ranges\
is only one order less than the increment $\Delta W$, see Eq. (\ref{3.11}).
Note that this behavior is inherent to all short pulses with large potential
steps belonging to exactly solvable cases, see Ref. \cite{AdGavGit17}.

As was already mentioned, among external fields of all exactly solvable
cases, only the external field of the new case, considered in the present
work, is given by an analytic function, which is not symmetric with respect
to its maximum value. This circumstance makes it possible to analyze
nonperturbatively exactly the influence of such asymmetry on effects of the
vacuum instability. In particular, the influence of the asymmetry on the
particle production can be seen by the example of considering the
differential numbers (\ref{3.2}). Due to this asymmetry these quantities
behave differently as functions of positive and negative longitudinal
momenta $p_{x}$. Apparently, a more detailed analysis of asymmetry effects
should be carried out by comparing the vacuum instability in the analytic
asymmetric field with the one in Sauter-like field, since the latter field
is similar but differs by the presence of an asymmetric right part. In this
case, it may be useful to clarify the role of the symmetry (with respect to
the middle point), which has no particular physical significance, inherent
in the configurations of external fields that are simplified for obtaining
exact solutions.

This is evident in the further analysis. For example, when analyzing the
differential mean numbers (see Eq. (\ref{3.14})) in the case of large $%
\sigma $, or equivalently in the case of the slowly varying analytic
asymmetric electric field, the function $\tau $ which determines the
behavior of these numbers, contains in the leading term in the denominator a
third-order term in longitudinal kinetic momenta at time $t_{\max } $, $%
p_{x}^{\prime }$ (see Eq. \ref{3.20}) which significantly changes the
behavior of the mean differential numbers over the regions of momenta in
comparison with their behavior (see Ref.\cite{GavGit96}) in the case of
symmetric Sauter-like electric field.

The exact results obtained make it possible to see clearly how the vacuum
instability behaves in a slowly varying field, which, in particular,
corresponds to large values of the parameter $\sigma $. In this case, it was
shown (see Eq. (\ref{3.20}) and the accompanying discussion) that, in a
fairly wide range of momenta $\pi _{\perp }\ll eE_{0}\sigma $, $%
p_{x}^{\prime }\ll eE_{0}\sigma $, the differential mean numbers do not
depend on $\sigma $\ and coincide with ones produced by a constant electric
field field, see Eq. (\ref{3.23}). It is also shown that the total mean
number of created pairs $N\approx V_{\left( d-1\right) }\rho ^{\mathrm{%
\Omega }}$, as well as the corresponding density $\rho ^{\mathrm{\Omega }}$,
are proportional to the increment $\Delta W$, as for other slowly varying
fields (see Ref. \cite{AdGavGit17}).

In Sec. \ref{S4}, we compare the total mean number $N$ of created pairs in
the regime of the slowly varying field with an estimate obtained in an
universal slowly varying field approximation proposed in Ref. \cite{GavGit17}%
, thus demonstrating the effectiveness of the latter. We stress that the
remarkable agreement with predictions of the universal slowly varying field
approximation is quite expected, since in this case\ the density $\rho ^{%
\mathrm{\Omega }}$\ is proportional to a large parameter, namely, the amount
of ``work'' that the field does by generating pairs. The shape of the
electric field determines the proportionality factor, see Eqs.\ (\ref{g3a})
and (\ref{3.41}). These factors are quite similar for strong electric
fields, but are significantly different otherwise. It also should be noted
that the cubic dependence of the differential mean numbers (\ref{3.20}) on
the longitudinal momenta $p_{x}^{\prime }$, which arises due to the
asymmetry, is essential for a correct calculation of the total mean number $N
$. Indeed, if for some reason, one omits the cubic term in representation (%
\ref{3.20}), the resulting total mean number $N$ would be significantly
different. In addition to the above, the field asymmetry affects the
differential mean numbers since the leading term in the factor $\tau ^{-1}$
depends on $p_{x}^{\prime 3}$. There was no such dependence for the case of
the symmetric Sauter-like field, as can be seen by comparing Eq. (\ref{3.20}%
) of the current work with the corresponding results in the article \cite%
{AdGavGit17}.

Finally, using the same approximation, we calculate the mean values of the
current density and the energy-momentum tensor of created particles.

\section{Acknowledgments}

The work is supported by Russian Science Foundation, grant No. 19-12-00042.

\section{Appendix. Some properties of hypergeometric functions\label{A1}}

The hypergeometric function $F\left( a,b,c;z\right) =\,_{2}F_{1}\left(
a,b,c;z\right) $ (here and in what follows it is supposed that parameters $a$
and $b$ are not equal to $0,-1,-2$, $\ldots $) is defined by series%
\begin{equation}
F\left( a,b,c;z\right) =\sum_{n=0}^{+\infty }\frac{\left( a\right)
_{n}\left( b\right) _{n}}{\left( c\right) _{n}}\frac{z^{n}}{n!}=\frac{\Gamma
\left( c\right) }{\Gamma \left( a\right) \Gamma \left( b\right) }%
\sum_{n=0}^{+\infty }\frac{\Gamma \left( a+n\right) \Gamma \left( b+n\right) 
}{\Gamma \left( c+n\right) }\frac{z^{n}}{n!},\ \ \left\vert z\right\vert <1.
\label{A01}
\end{equation}%
Note that in the solutions (\ref{2.13}) and (\ref{2.14}) the arguments $%
1-\xi ^{-1}$ and $\xi ^{-1}$ in the corresponding hypergeometric functions
are less than unity and the series (\ref{A01}) converges.

At $\left\vert z\right\vert =1$ the series (\ref{A01}) converges absolutely
when $\mathrm{Re}\left( c-a-b\right) >0$. The integral representation%
\begin{equation}
F\left( a,b,c;z\right) =\frac{\Gamma \left( c\right) }{\Gamma \left(
b\right) \Gamma \left( c-b\right) }\int_{0}^{1}t^{b-1}\left( 1-t\right)
^{c-b-1}\left( 1-zt\right) ^{-a}dt,\ \ \left( \mathrm{Re }c>\mathrm{Re }
b>0\right)  \label{A02}
\end{equation}%
gives an analytical continuation for the function $F\left( a,b,c;z\right) $
to the complex $z$-plane with a cut along the real axis from $1$ to $\infty $
(since the right-hand side is an unambiguous analytic function in the domain 
$\left\vert \arg \left( 1-z\right) \right\vert \leq \pi $). From the
integral representation (\ref{A02}) it is easy to see that $%
\lim_{z\rightarrow 0}F\left( a,b,c;z\right) =1.$ The formula for
differentiating the hypergeometric function has the form: 
\begin{equation}
\frac{d}{dz}F\left( a,b,c;z\right) =\frac{ab}{c}F\left( a+1,b+1,c+1;z\right)
.  \label{AdF}
\end{equation}

It is follows from (\ref{A02}) that 
\begin{equation}
F\left( a,b,c;z\right) =\left( 1-z\right) ^{c-a-b}F\left( c-a,c-b,c;z\right)
,\ \ \left\vert z\right\vert <1.  \label{A04}
\end{equation}

Hypergeometric function can be transformed as 
\begin{eqnarray}
&&\ F\left( a,b,c;z\right) =\frac{\Gamma \left( c\right) \Gamma \left(
c-a-b\right) }{\Gamma \left( c-a\right) \Gamma \left( c-b\right) }F\left(
a,b,a+b-c+1;1-z\right)  \notag \\
&&+\left( 1-z\right) ^{c-a-b}\frac{\Gamma \left( c\right) \Gamma \left(
a+b-c\right) }{\Gamma \left( a\right) \Gamma \left( b\right) }F\left(
c-a,c-b,c-a-b+1;1-z\right) ,\ \ \left( \left\vert \arg \left( 1-z\right)
\right\vert <\pi \right) ,  \label{A05a}
\end{eqnarray}

The hypergeometric equation in its general form, 
\begin{equation}
z\left( 1-z\right) w^{\prime \prime }\left( z\right) +\left[ c-\left(
a+b+1\right) z\right] w^{\prime }\left( z\right) -abw\left( z\right) =0,
\label{A06}
\end{equation}%
has three regular singular points $z=0,1,\infty $. When none of the numbers $%
c$, $c-a-b$, $a-b$ is integer, the general solution $w\left( z\right) $ of
the hypergeometric equation (\ref{A06})\ can be obtained as 
\begin{eqnarray}
&&w\left( z\right) =c_{1}w_{1}\left( z\right) +c_{2}w_{2}\left( z\right) ,\ 
\text{ }z\rightarrow 0,  \notag \\
&&w\left( z\right) =c_{1}w_{3}\left( z\right) +c_{2}w_{4}\left( z\right) ,\
\ z\rightarrow 1,  \notag \\
&&w\left( z\right) =c_{1}w_{5}\left( z\right) +c_{2}w_{6}\left( z\right) ,\
\ z\rightarrow \infty \text{.}  \label{A07set}
\end{eqnarray}%
where $c_{1}$ and $c_{2}$ are some constants, and the functions $w_{j}\left(
z\right) $, $j=1,\ldots ,6$, have the form: 
\begin{eqnarray}
&&w_{1}\left( z\right) =F\left( a,b,c;z\right) ,\ \ w_{2}\left( z\right)
=z^{1-c}F\left( a-c+1,b-c+1,2-c;z\right) ,  \notag \\
&&w_{3}\left( z\right) =F\left( a,b,a+b+1-c;1-z\right) ,\   \notag \\
&&w_{4}\left( z\right) =\left( 1-z\right) ^{c-a-b}F\left(
c-b,c-a,c-a-b+1,1-z\right) ,  \notag \\
&&w_{5}\left( z\right) =z^{-a}F\left( a,a-c+1,a-b+1,z^{-1}\right) ,\   \notag
\\
&&w_{6}\left( z\right) =z^{-b}F\left( b,b-c+1,b-a+1,z^{-1}\right) .
\label{A08}
\end{eqnarray}%
The Kummer relations and for the hypergeometric equation \cite{Bateman}
allow us to represent the functions $w_{1}\left( z\right) $ and $w_{2}\left(
z\right) $ via the functions $w_{3}\left( z\right) $ and $w_{4}\left(
z\right) $,%
\begin{eqnarray}
&&w_{1}\left( z\right) =e^{i\pi \left( 2\alpha _{1}-b\right) }\frac{\Gamma
\left( 2\left( \alpha _{1}+1\right) -a-b\right) \Gamma \left( b-a+1\right) }{%
\Gamma \left( 2-a\right) \Gamma \left( 2\alpha _{1}-a+1\right) }w_{4}\left(
z\right)  \notag \\
&&-e^{i\pi \left( 2\alpha _{1}-a\right) }\frac{\Gamma \left( 2\left( \alpha
_{1}+1\right) -a-b\right) \Gamma \left( a-b-1\right) }{\Gamma \left(
1-b\right) \Gamma \left( 2\alpha _{1}-b\right) }w_{3}\left( z\right) ,
\label{cm2} \\
&&w_{2}\left( z\right) =e^{i\pi \left( a-1\right) }\frac{\Gamma \left(
a+b-2\alpha _{1}\right) \Gamma \left( b-a+1\right) }{\Gamma \left( b-2\alpha
_{1}+1\right) \Gamma \left( b\right) }w_{4}\left( z\right)  \notag \\
&&+e^{i\pi b}\frac{\Gamma \left( a+b-2\alpha _{1}\right) \Gamma \left(
a-b-1\right) }{\Gamma \left( a-2\alpha _{1}\right) \Gamma \left( a-1\right) }%
w_{3}\left( z\right) .  \label{cm1}
\end{eqnarray}

\end{document}